\documentclass[useAMS,usenatbib]{mn2e}
\usepackage{graphicx}
 \usepackage{times}
 \usepackage{rotating}
\usepackage{txfonts}

\def\va{\hbox{OGLE 05155332-6925581}}

\DeclareGraphicsExtensions{.gif,.pdf,.bmp,.jpg, .eps}

 \def\gtrsim{\mathrel{\hbox{\rlap{\hbox{\lower4pt\hbox{$\sim$}}}\hbox{$>$}}}}
 \def\ltsim{\mathrel{\hbox{\rlap{\hbox{\lower4pt\hbox{$\sim$}}}\hbox{$<$}}}}

\title[Physical parameters and evolutionary route of OGLE 05155332-6925581]{Physical parameters and evolutionary route for the  LMC  interacting binary OGLE 05155332-6925581\thanks{Based on observations carried out at ESO telescope: ESO proposal 082.D-0575(A).}}

\author[H. E. Garrido et al.]
  {H. E. Garrido$^{1,2}$\thanks{E-mail: hgarrido@eso.org},
R. E. Mennickent$^{2}$, G. Djura{\v s}evi\'c$^{3,4}$, Z. Ko{\l}aczkowski$^{5}$,     E. Niemczura$^{5}$  \newauthor  and N. Mennekens$^{6}$ \\
       $^{1}$ European Organisation for Astronomical Research in the Southern Hemisphere, Alonso de Cordoba 3107, Vitacura, Casilla 19001, Santiago 19, Chile\\
        $^2$Universidad de Concepci\'on, Departamento de Astronom\'{\i}a,
      Casilla 160-C, Concepci\'on, Chile\\
  $^3$ Astronomical Observatory, Volgina 7, 11060 Belgrade, Serbia\\
  $^{4}$ Isaac Newton Institute of Chile, Yugoslavia Branch, 11000 Belgrade, Serbia\\
  $^{5}$ Instytut Astronomiczny Uniwersytetu Wroc{\l }awskiego Ul. Kopernika 11, 51-622 Wroc{\l }aw, Poland\\
    $^{6}$ Astrophysical Institute, Vrije Universiteit Brussel, Pleinlaan 2, 1050 Brussels, Belgium}
\date{}

\pagerange{\pageref{firstpage}--\pageref{lastpage}} \pubyear{2012}

\def\LaTeX{L\kern-.36em\raise.3ex\hbox{a}\kern-.15em
    T\kern-.1667em\lower.7ex\hbox{E}\kern-.125emX}

\begin{document}

\label{firstpage}

\maketitle

\begin{abstract}
{We analyze multicolor light curves and high resolution optical spectroscopy of the eclipsing binary and Double Periodic Variable OGLE 05155332-6925581.
According to Mennickent et al., this system shows a significant change in the  long non-orbital photometric cycle, a loop in the color-magnitude diagram during this cycle and
discrete spectral absorption components that were interpreted as evidence of systemic mass loss.  We find
that the best fit to the multi-band light curves requires a circumprimary optically thick disc with a radius about twice the radius of the more massive star.
The spectroscopy indicates a mass ratio of $0.21\pm 0.02$ and masses for the hot and cool stars of $9.1\pm 0.5$ and $1.9\pm0.2$ M$_{\odot}$, respectively.
A comparison with synthetic  binary-star evolutionary models indicates that the system has an age of 4.76 $\times$ 10$^{7}$ years,
is in the phase of rapid mass transfer, the second one in the life of this binary, in a Case-B mass-exchange stage.
Donor-subtracted H$\alpha$ profiles show the presence of double emission formed probably in an optically thin circumstellar medium, while the variable He\,I profile and the H$\beta$ absorption wings are probably formed in the optically thick
circumprimary disc.
The model that best fit the observations shows  the system with  a relatively large mass transfer rate of $\dot{M} = 3.1\times 10^{-6}$ M$_{\odot}$/yr. However, the orbital period remains relatively stable during
almost 15 years. This observation suggests that the hot-spot mass-loss model proposed by other authors is not adequate in this case, and that some other mechanism is efficiently removing angular momentum from the binary. Furthermore, our observations suggest that the DPV phenomenon could have an important effect in the balance of mass and angular momentum in the system. }

\end{abstract}

\begin{keywords}
stars: early-type-stars: evolution- stars: mass-loss,-stars: emission-line, binaries:eclipsing.
stars: variables:others
\end{keywords}

\section{Introduction}
\cite{M1} (hereafter, M08), observed spectroscopically the  eclipsing LMC  star \va (MACHO IDs 79.5739.807 and 78.5739.78; OGLE -SC8-125836). This star is a member of the group of close binaries named Double Periodic Variables (DPVs; Mennickent et al. 2003, 2005a). These stars show two closely related photometric periodicities; one longer and not exactly cyclic (Mennickent et al. 2005b) and one shorter that reflects the binary period.

The investigation carried out by M08 was based on  multi-wavelength light curve analysis, and resulted in an intermediate-mass semi-detached Algol-type binary in an evolutionary stage characterized by mass exchange and mass loss  where the less massive star transfers matter onto the more massive  star. M08 also  detected evidence for a relatively luminous disc around the primary star. They reported a loop in the color-magnitude diagram  during the long cycle that was interpreted as  evidence of mass loss modulated by some still unknown mechanism. This was also supported by the motion of discrete absorption components observed in the infrared hydrogen lines. M08 offered the explanation of cyclic mass loss for the long photometric cycle, but failed in identifying the motion of the gainer in their low resolution spectra, and the stellar parameters, in particular the mass ratio, remained relatively uncertain in their study.

Consequently, we decided to investigate this system with high resolution spectroscopy, in order to look for the missing features of the hot star, find definitive stellar parameters and understand better the evolutionary stage of the system when comparing with results of  non-conservative evolutionary models for binary stars.
 In this paper  we present our study of this high-resolution spectroscopy.
In the next section we give  a detailed report of the observations used for the analysis; our results are presented in Section 3; a detailed discussion of these results is given in Section 4 and our conclusions are presented in Section 5.


\section{ observations and data reduction}

We have monitored \va\  spectroscopically in the optical wavelength range. The observations of this object were obtained in the period  from 17 January to 26 September of 2009, using the Fibre Large Array Multi-Element Spectrograph (Flames) on the Very large Telescope (VLT) in service mode.  In total 19 spectra were obtained, 14 of them with the Giraffe spectrograph using the standard Medusa-Giraffe setting HR5A in the wavelength region of $4\,341$\,-\,4\,585 $\AA$ and 5 spectra using the UVES spectrograph in the region 4\,775\,-\, 6\,800 $\AA$  with standard UVES setting 580, as shown in Table 1. The UVES data do not include the region $5\,756$ to $5\,823$ $\AA$ because of the gap between the two detectors.

\begin{table}\label{Table_1}
\caption{Summary of the wavelength coverage, exposure time and spectral resolution of the different FLAMES modes and settings used in the observation of \va. $R$ is the spectral resolving power and $N$ the number of spectra obtained.}
\begin{tabular}{llcccc}
\hline
Mode & Setting &$\lambda-$range (\AA)  & Exp. Time (s) & R & N.\\\hline
Medusa & HR5A &  4341-4585 & 2770 & 20000& 14\\
UVES & 580 & 4775-6800 &2770 & 47000&5\\
\hline
\end{tabular}

\end{table}

The ESO Common Pipeline Library (CPL) FLAMES reduction routines were used for initial data processing and consisted  of flat-fielding, bias subtraction and wavelength calibration. Separate fibers were used to observe the sky in each exposure, each sky fibre was inspected for signs of  cross-contamination from bright spectra/emission lines from adjacent fibers on the detector. Any contaminated sky fiber was rejected before creating a median sky spectrum, which was then subtracted from each science image. Related to the sky subtraction, one of the principal limitations of the fiber spectroscopy is the subtraction of local nebular emission. Due to the distance of the Magellanic Clouds, even long-slit spectroscopy can suffer difficulties from spatially-varying nebular emission (\cite{W1}). Because we selected Balmer and He I lines for our analysis, we examined these lines for residuals of nebulosity subtraction. These were actually found near the H$\alpha$ profile, as shown later, but the affected regions are easily identified and isolated, and not considered in our  analysis.  The above method was implemented by \cite{E1} to reduce  VLT-Flames observations.

The UVES frames are reduced using a similar method as for the Giraffe data. We removed cosmic rays and then extracted the spectra, that were normalized over the entire spectral region. All spectra discussed in this paper are  corrected to the heliocentric frame, using the RVCORRECT and DOPCOR packages implemented in IRAF\footnote{IRAF is distributed by the National Optical Astronomy Observatories, which are operated by the Association of Universities for Research in Astronomy, Inc., under cooperative agreement with the National Science Foundation.}. We use indifferently the words primary (\textit{gainer}) and secondary (\textit{donor}) for the components of the binary system as usual in the literature of mass exchanging semi-detached binaries.

 \begin{table*}\label{Table_2}
\caption{RVs measured from the Medusa and UVES  spectra for \va  referred to the Local Standard of Rest. The orbital phase $\Phi_{o}$ is calculated following the ephemeris given in Eq.\,(1). Subindex 1 and 2 refers to the gainer and donor, respectively.}
\begin{tabular}{lclclcll}
\hline
HJD & $\phi_{o}$ & RV$_{1}$ & (O-C)$_{1}$ & RV$_{2}$ & (O-C)$_{2}$ & Mode\\
-2450000& & km s$^{-1}$ &  km s$^{-1}$ & km s$^{-1}$ & km s$^{-1}$& \\\hline
5084.8895 & 0.043 & $228.82 \pm 2.63$& $-12.32 \pm 2.63$& $305.71\pm 1.18\,^{b}$ & $-0.61\pm 1.18$ &Medusa \\
5100.8791 & 0.238 & $187.48\pm 1.68$ &$ -8.44\pm 1.68$ & $455.29\pm 1.37$ &$ 2.28 \pm 1.37$ & Medusa \\
4853.6087 & 0.292 & $178.59\pm0.79 \,^{a}$ & $-11.21\pm0.79$ & & &UVES \\
4853.6538 & 0.298 & $187.94\pm 1.50$ & $-1.45 \pm 1.50$&$ 444.38\pm 0.75\,^{b}$ &$ -0.14\pm 0.75$ & Medusa \\
4868.6050 & 0.351 & $197.64\pm0.46  \,^{a}$ & $9.30\pm0.46 $ & & &UVES\\
5079.8642 & 0.353 & $185.83\pm0.94$ & $-2.55\pm0.94$ & $407.31\pm 0.48\,^{b} $& $-5.67\pm 0.48$ & Medusa \\
4868.6495 & 0.357 & $183.77\pm1.24 $ & $-4.73\pm1.24 $ & $417.92\pm 1.50\,^{b}$ & $8.04\pm 1.50$ & Medusa \\
4876.5457 & 0.441 & $215.14\pm1.89 \,^{a}$ & $18.33\pm1.89$ & & & UVES\\
4898.6029 & 0.469 & $207.32\pm 0.51 $ $^{a}$ & $5.56 \pm 0.51 $& & &UVES \\
5080.8272 & 0.485 & $ 215.98\pm 1.54$ &$ 10.98 \pm 1.54$& $265.49\pm 1.02 \,^{b}$ & $-6.14\pm 1.02$ & Medusa \\
4848.5946 & 0.604 & $228.10\pm 0.98$ & $-6.29 \pm 0.98$ & $133.30 \pm 0.34 $ & $2.69\pm 0.34 $ & Medusa \\
4848.6366 & 0.610 & $245.81\pm 1.25  \,^{a}$ &$ 9.87\pm 1.25$ & & &UVES\\
5081.8340 & 0.623 & $223.96 \pm 1.41$ & $-15.33  \pm 1.41$& $116.96\pm 0.72$ & $4.48\pm 0.72$ & Medusa \\
4892.5550 & 0.639 & $ 233.16\pm 0.98$ & $-10.16\pm 0.98$ & $100.81\pm1.21 \,^{b}$ &$ 2.06\pm1.21$ &Medusa\\
4863.5947 & 0.663 & $238.02\pm3.16 $ &$ -11.07\pm3.16 $ & $80.23 \pm 1.33$ & $-0.88\pm 1.33$ & Medusa \\
4907.5324 & 0.695 & $261.07\pm1.81$ & $4.98\pm1.81$ & $61.32\pm 0.16\,^{b}$ &$ -2.39\pm 0.16$ & Medusa \\
5082.8394 & 0.761 & $264.24\pm1.80$ & $-2.68\pm1.80$ & $56.16\pm0.60\,^{b}$ &$ 3.85\pm0.60$ & Medusa\\
4865.5906 & 0.937 & $280.54\pm2.76$ & $17.01\pm2.76$ & $180.51 \pm 0.88$ & $5.29 \pm 0.88$ & Medusa \\
5076.9004 & 0.946 &$ 272.10\pm 1.73$ & $10.02\pm 1.73$& $180.56\pm 1.58\,^{b}$ &$ -5.26\pm 1.58$ & Medusa \\\hline
\end{tabular}
\vspace{0.2 cm}
\noindent\parbox[c]{17.3 cm}{ \small{Note. $^{a}$ RVs values based on the line He II 4\,921 \AA , $^{b}$ RVs values based on the lines Mg II 4\,481 \AA, only.} }
\end{table*}

\section{ Results}

\subsection{ Radial velocity measurements and spectroscopic mass-ratio}\label{spectroscopy}

M08 could not estimate a spectroscopic mass ratio for this system, mainly
due to the weakness of the  features of the primary star  in the available low resolution spectra.
In addition, the analysis of the OGLE II and MACHO light curves provided inconsistent values  for the photometric mass ratio,
in part since the light curve model at that time did not include a contribution from a potential circumprimary disc. In this section we provide the first detection of the features of the hotter component and  an improved determination for the the system mass ratio.

Line identification was done with the aid of the spectral atlas for  B main-sequence stars
provided  on-line  by Heidelberg University\footnote{http://www.lsw.uni-heidelberg.de/cgi-bin/websynspec.cgi}.  A list of the most prominent lines identified in the spectra are given in Table 3.
We identified several strong absorption lines of He\,I, Mg\,II and  Fe\,II  in our Flames spectra.
The UVES spectra display mainly Balmer and strong He\,I lines.
H$\alpha$ profiles are  broad absorptions filled by emission. In this line we detected  contributions of both stellar components plus a very narrow central absorption spike. This feature turned to be residual absorption from the process of sky subtraction, consistent with spatially variable nebular emission in this part of the sky.
The  He\,I  lines of the primary  appear to be partially blended with lines from the secondary star, indicating
a relatively hot donor star.

\begin{table}\label{Table_2.1}
\caption{Principal lines detected in the optical spectrum of \va. The symbol $\dagger$ indicate lines used in the RV measured.}
\begin{tabular}{lcclcclc}
\hline
$\lambda\, (\AA)$ &  Line & & $\lambda\, (\AA)$  &  Line & & $\lambda\, (\AA)$ & Line \\\hline
4\,387.9 & He I &    & 4\,471.5$\dagger$  & He I  & &  4\,390.6 &Mg II  \\
4\,481.2$\dagger$  & Mg II & &  4\,549.5 $\dagger$ & Fe II & & 4\,861.3  & H$\beta$ \\
4\,921.9$\dagger$ & He I & & 5\,015.7& He I & & 5\,875.6  & He I \\
5889.9& Na I int.  & &5\,895.2  & Na I int. & & 6\,562.8  & H$\alpha$ \\
 6\,678.2 & He I && & &&&\\
\hline
\end{tabular}

\end{table}

\begin{figure}
    \centering
    \includegraphics[width=0.53\textwidth, height=0.5\textwidth]{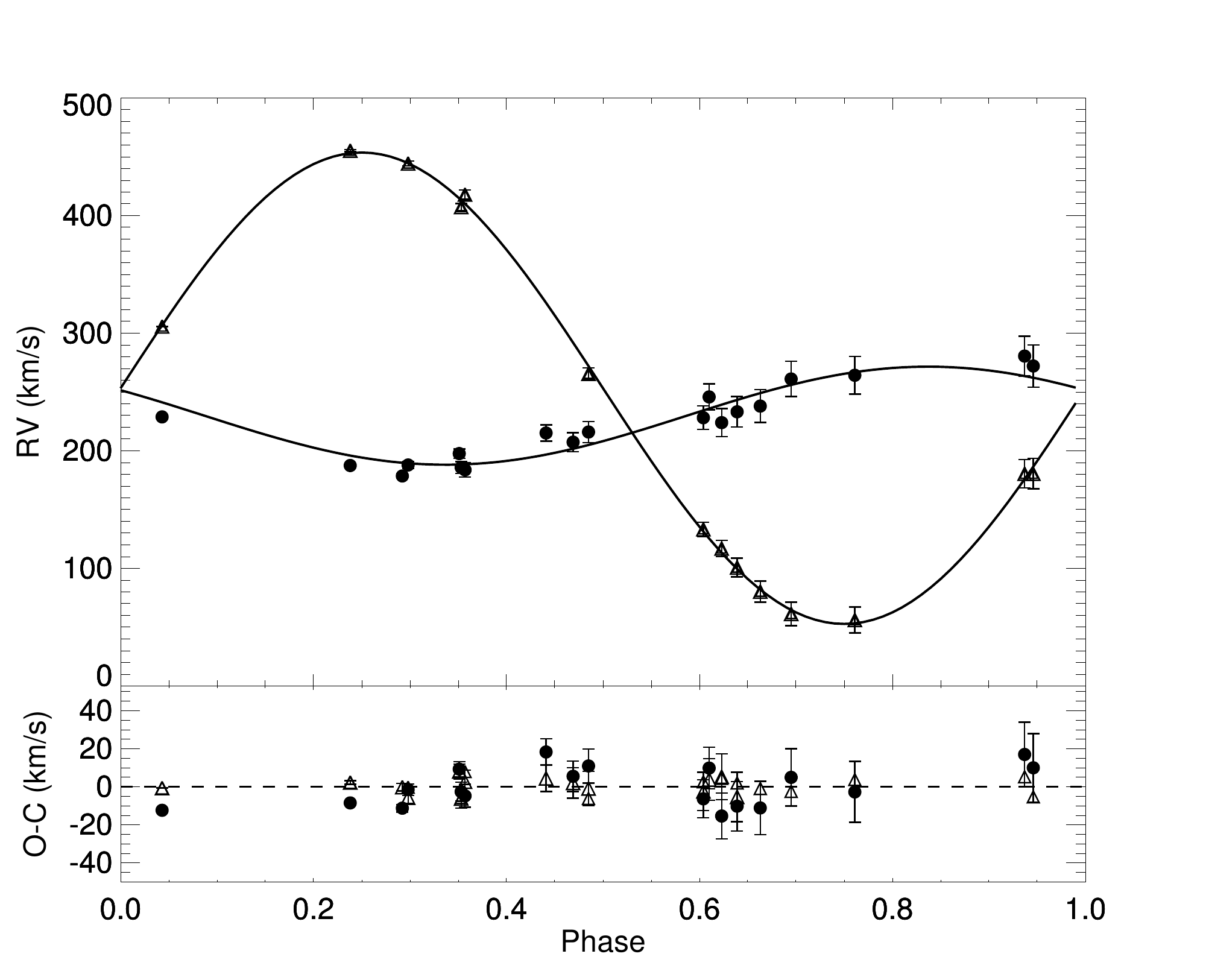}
    \caption{The RV curve together with O-C for \va. Plot is folded to orbital period using the ephemeris for the main minimum, given in the equation \ref{ephemeris}. Panel shows the RV measurements with the best-fit solutions. Filled circles represent the gainer and triangles the donor. The bottom panel shows the residuals.}
    \label{RV}
\end{figure}

All spectra were normalized to the continuum before performing the RV measurements and analysis. The RVs were
measured by calculating the positions of the line centre  with a gaussian fit using the IRAF SPLOT task.
We measured the heliocentric radial velocities for the secondary using the Mg\,II\,4\,481\AA\  and Fe\,II\,4\,549 \AA\  lines.
These RVs are listed in Table \ref{Table_2}, along with the corresponding orbital phase computed with the ephemerides given in Section \ref{light_curves_analysis}  and the errors taken as  the standard deviations of the measurements. All velocities are considered with equal statistical weight. We assumed a circular orbit, so a sine function was selected to fit them, obtaining  a systemic velocity $\gamma_{\mathrm{donor}}=253.20 \pm 1.22 \  \mathrm{km\, s}^{-1}$ and semi-amplitude $K_{\mathrm{donor}}= 200.42 \pm 1.66 \  \mathrm{km\, s}^{-1}$ with rms 4 $\mathrm{km\, s}^{-1}$ (see Fig. \ref{RV}).

The lines He\,I\,4\,471\,\AA\  and He\,I\,4\,921\,\AA\   mainly follow the primary star, but a minor contribution from the secondary star is present. We used a deblending method  to isolate the components in the analysis. We performed this task with the IRAF deblending 'd' routine available in the SPLOT package. We  carefully measured the RVs of the He\,I lines (see Table \ref{Table_2}) and fit them with a sine function, obtaining  a systemic velocity
 $\gamma_{\mathrm{gainer}}= 229.85 \pm 2.87\  \mathrm{km\, s}^{-1}$ and  a semi-amplitude $K_{\mathrm{gainer}}= -41.69 \pm 3.67 \  \mathrm{km\, s}^{-1}$ with rms 11 $\mathrm{km\, s}^{-1}$.
Although there are other He\,I and Balmer lines in  our UVES spectra with similar double  profiles around phases 0.29 and 0.61, we could not de-blend them to include them in the radial velocity calculation, since the individual components were hardly  resolved. The difference in the systemic velocity between the solution for the primary and secondary  can be explained by emission of circumstellar matter affecting the He\,I lines.
From the above semi-amplitudes we find a spectroscopic mass ratio of  $q_{sp}= 0.21\pm 0.02$ for \va, which is smaller than  $q$ = 0.29 obtained by M08  from the best MACHO blue light-curve fitting model.

Interestingly, our donor half-amplitudes derived from the study of Fe\,II  and Mg\,II lines are larger than those found by M08 in the Balmer lines by up to 80 km\,s$^{-1}$. This could be associated to a  larger contamination by emission in Balmer lines, as suggested by our donor-subtracted H$\alpha$ profiles (Section 3.3). The fact that the emission formed around the gainer moves in anti-phase with the donor absorption produces the net effect of lowering the observed donor radial velocity half-amplitude. We considered another possible cause for the half-amplitude discrepancy; displacement of the light-centre due to donor irradiation by the disc and gainer (e.g. Wade \& Horne 1988, Watson et al. 2003). This effect would result in an over-estimate of $K_{2}$, leading to a corresponding under-estimate of $q$. However,
 the absence of asymmetries in the donor line profiles and the large observed differences  in H\,I line radial velocity half-amplitudes (up to 45 km\,s$^{-1}$), suggest that
residual emission is the most likely explanation. It is difficult to imagine how light-centres could differ so much between lines of the same element at a given ionization stage.

  \begin{figure*}
    \centering
    \includegraphics[]{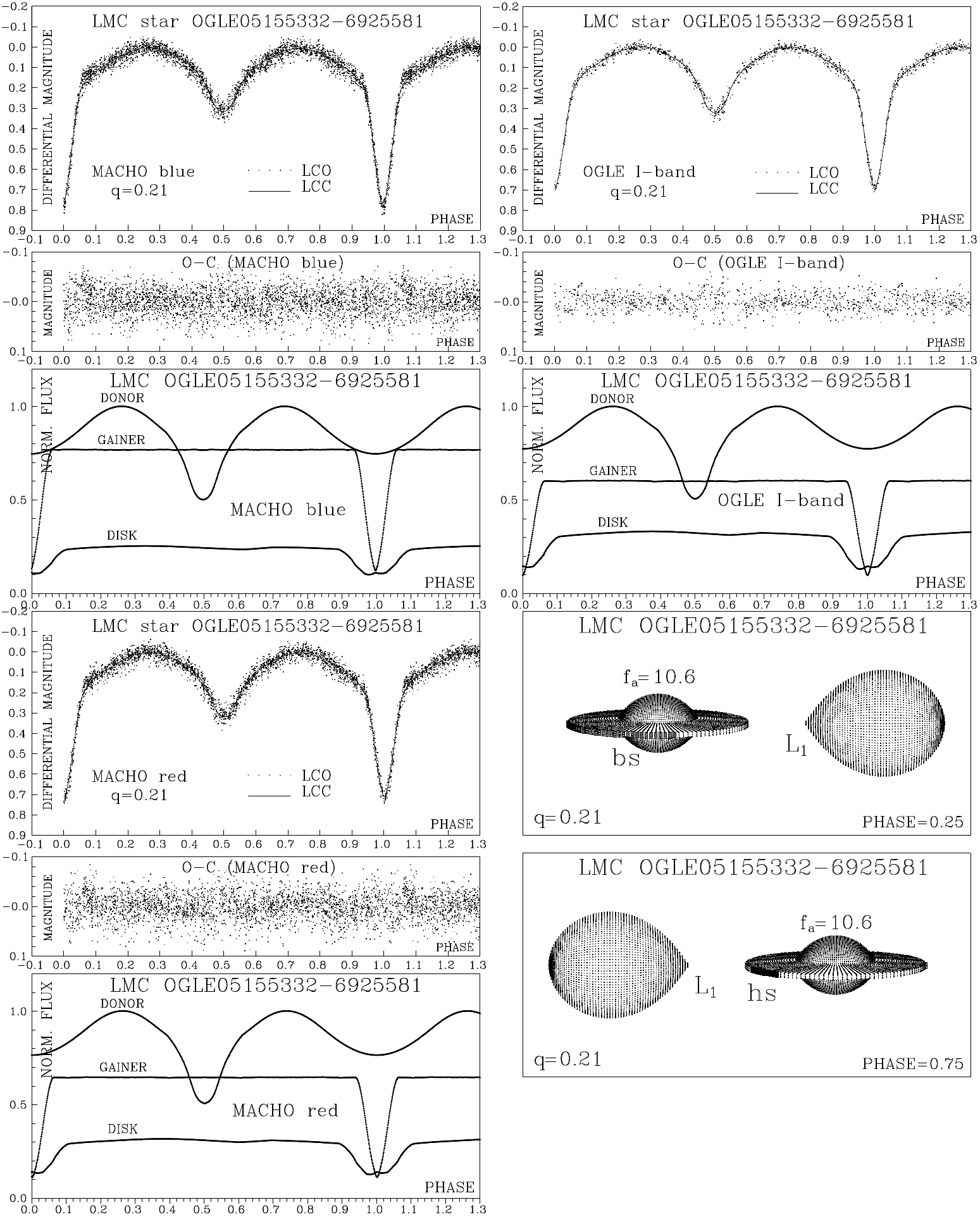}

    \caption{Observed (LCO) and synthetic (LCC) light-curves of
LMC star OGLE 05155332-6925581 obtained by analyzing MACHO blue, red and OGLE I-band photometric observations; final
O-C residuals between the observed and optimum synthetic light curves;
fluxes of donor, gainer and of the accretion disc, normalized
to the donor flux at phase 0.25; the views
of the optimal model at orbital phases 0.25 and 0.75,
obtained with parameters estimated by the light curve analysis.}
    \label{fOGLE-n-synch}
\end{figure*}

 \begin{table*}\label{Table_3}
\caption{Result of the analysis of LMC star OGLE $05155332-6925581$ in MACHO blue, MACHO red and OGLE I-band light curves obtained by solving the inverse problem for the Roche model with an accretion disc around the more-massive (gainer) component in critical rotation regime.}
\begin{tabular}{lllllll}
\hline
Quantity & MACHO blue & MACHO red & OGLE I & mean BRI & Quantity & \\\hline
$n$ & 2\,964 & 2\,796 & 883 &    & ${\cal M}_{h}[{\cal M}_{\odot}]$ & $9.1\pm0.5$\\
$\Sigma (O-C)^{2}$ & 1.9490 & 1.8550 & 0.2634&  & ${\cal M}_{c}[{\cal M}_{\odot}]$ & $1.9\pm 0.2$\\
$\sigma_{\mathrm{rms}}$ & 0.0256 & 0.0258 & 0.0173      &       &   ${\cal R}_{h}[ \mathrm{R}_{\odot}]$ & $5.6\pm 0.2$\\
$i[^{\circ}]$ & 82.33 & 82.25 & 82.36 & $82.3\pm 0.3$ & ${\cal R}_{c}[\mathrm{R}_{\odot}]$ & $8.9\pm 0.3$\\
F$_{\mathrm{d}}$ & 0.76 & 0.77 & 0.77 & $0.77\pm 0.03$ & $\log g_{h}$ & $3.9\pm 0.1$  \\
T$_{d}[\mathrm{K}]$ & 12\,200 & 12\,800 & 12\,770 & $12\,600\pm 600$ & $\log g_{c}$ & $2.8\pm 0.1$  \\
d$_{\mathrm{e}}[a_{\mathrm{orb}}]$ & 0.020 & 0.021 & 0.025 & $0.02\pm 0.01$& $M^{\mathrm{h}}_{\mathrm{bol}}$ & $-5.3\pm 0.2$\\
d$_{\mathrm{c}}[a_{\mathrm{orb}}]$ & 0.070 & 0.072 & 0.077 & $0.07\pm 0.01$ & $M^{\mathrm{c}}_{\mathrm{bol}}$ & $-3.4\pm 0.1$\\
a$_{\mathrm{T}}$ &4.0 & 3.5 & 3.5 & $3.7\pm 0.3$ & $a_{\mathrm{orb}}[\mathrm{R}_{\odot}]$ & $35.2\pm 0.5$\\
$\mathrm{f}_{\mathrm{h}}$ & 10.6 & 10.6 & 10.5 & $10.6\pm 0.3$ & ${\cal R}_{d}[\mathrm{R}_{\odot}]$ &   $14.1\pm 0.5$\\
F$_{\mathrm{h}}$ & 1.00 & 1.00 & 1.00 & 1.00 & d$_{\mathrm{e}}[\mathrm{R}_{\odot}]$ & $0.8\pm 0.3$\\
T$_{\mathrm{c}}$[K] & 12\,930 & 12840 & 12\,830 & $12\,900\pm 500$ &  d$_{\mathrm{c}}[\mathrm{R}_{\odot}]$ & $2.6\pm 0.5$\\
A$_{\mathrm{hs}}=$T$_{\mathrm{hs}}\,/\,$T$_{d}$ & 1.13 & 1.12 & 1.10 & $1.12\pm 0.1$ & & \\
$\theta_{\mathrm{hs}}[^{\circ}]$ & 14.1 & 16.3 & 16.2 & $15.5\pm 1.0$ & &\\
$\lambda_{\mathrm{hs}}[^{\circ}]$ & 327.4 & 322.9 & 322.4 & $324.2\pm 5.0$  & & \\
$\theta_{\mathrm{rad}}[^{\circ}]$ & -34.2 & -33.8 & -34.9 & $-34.3 \pm 5.0$ & & \\
A$_{\mathrm{bs}}=$T$_{\mathrm{bs}}\,/\,$T$_{d}$ & 1.15 & 1.19 & 1.20 & $1.18\pm 0.1$  & & \\
$\theta_{\mathrm{bs}}[^{\circ}]$ & 54.1 & 52.2 & 48.2 & $51.5\pm 3.0$ & & \\
$\lambda_{\mathrm{bs}}[^{\circ}]$ & 120.5 & 136.5 & 134.2 & $130.4\pm 13.0$ & & \\
$\Omega_{\mathrm{h}}$ & 7.94 & 7.92 & 7.90 & $7.92\pm 0.02$ & & \\
$\Omega_{\mathrm{c}}$ & 2.56 & 2.56 & 2.56 & $2.56 \pm 0.02$ & & \\\hline
\end{tabular} \\
\vspace{0.2 cm}
\noindent\parbox[c]{17.3 cm}{ \small{FIXED PARAMETER: $q={\cal M}_{c}/{\cal M}_{h}=0.21$ - mass ratio of the components, T$_{\mathrm{h}}=25\,000$ K - temperature of the more-massive (gainer), F$_{\mathrm{c}}=1.0$ -  filling factor for the critical Roche lobe of the donor,  F$_{\mathrm{h}}=\mathrm{R}_{h}/\mathrm{R}_{zc}=1.0$ -filing factor for the critical non-synchronous lobe of the more massive gainer (ratio of the stellar polar radius to the critical non-synchronous lobe radius along z-axis for a star in critical rotation regime), $f_{c}=1.00$ -synchronous rotation coefficients of the donor, $\beta_{\mathrm{h,c}}=0.25$ - gravity-darkening coefficients of the components,  A$_{\mathrm{h,c}}=1.0$ - albedo coefficients of the components. \\
Note: $n$ - total number of observations in MACHO blue (B), red (R) and OGLE I-band; $\Sigma (O-C)^{2}$ - final sum of square of residual between observed (LCO) and synthetic (LCC) light-curves, $\sigma_{\mathrm{rms}}$ - root-mean-square of the residuals, $i$ - orbit inclination (in arc degrees), F$_{\mathrm{d}}=\mathrm{R_{d}}/\mathrm{R_{yc}}$ - disc dimension factor (the ratio of the disc radius to the critical Roche lobe radius along y-axis), T$_{\mathrm{d}}$ - disc-edge temperature , d$_{\mathrm{e}}$, d$_{\mathrm{c}}$ - disc thicknesses (at the edge and at the centre of the disc, respectively) in the units of the distance between the components, a$_{\mathrm{T}}$ -disc temperature distribution coefficient,  f$_{\mathrm{h}}$ - non-synchronous rotation coefficient of the more massive gainer (in the critical rotation regime), T$_{\mathrm{c}}$ -temperature of the less massive cooler donor,  $\Omega_{\mathrm{h, c}}$ - dimensionless surface potentials of the hotter gainer and cooler donor, ${\cal M}_{h, c}[{\cal M}_{\odot}]$, ${\cal R}_{h, c}[\mathrm{R}_{\odot}]$ - stellar masses and mean radii of star in solar units, $\log g_{h, c}$ - logarithm (base 10) of the system components effective gravity,  $M^{\mathrm{h, c}}_{\mathrm{bol}}$ - absolute stellar bolometric magnitudes,  $a_{\mathrm{orb}}[\mathrm{R}_{\odot}]$,  ${\cal R}_{d}[\mathrm{R}_{\odot}]$,  d$_{\mathrm{e}}[\mathrm{R}_{\odot}]$, d$_{\mathrm{c}}[\mathrm{R}_{\odot}]$ - orbital semi-major axis, disc radius and disc thicknesses at its edge and centre, respectively, given in solar units. } }
\end{table*}

\subsection{The light curve analysis and model of the accretion disc}\label{light_curves_analysis}

We have used public domain photometric data available from the MACHO (\cite{A1}) and OGLE  II and III databases (\cite{U1}, \cite{S1} and \cite{P1}).
The photometric data  cover 13.7 years. The observations were folded to the orbital period using the ephemeris for the main minimum from M08:

\begin{equation}\label{ephemeris}
\mathrm{T_{min}(HJD)}=2\,450\,000.1392\,(21)+7^{\mathrm{d}}.284297\, (10)\times\mathrm{E}.
\end{equation}

\noindent The light-curve analysis was performed using the inverse-problem solving method (\cite{D1}) based on the simplex algorithm and the current version of the model of a binary system with a circumprimary disc (\cite{D2}). The model and code have been widely used and tested
during our recent research of intermediate-mass interacting binaries (e.g. Djura{\v s}evic et al. 2010, 2011, 2012, Mennickent et al. 2012a).
For our calculations we assume that the disc  is optically thick and its edge can be approximated by a cylindrical surface. The thickness of the disc can change linearly with radial distance, allowing the disc to take a conical shape (that it can be convex, concave or plane-parallel). The geometrical properties of the disc are determined by its radius (${\cal R}_{d}$), its thickness at the edge (d$_{\mathrm{e}}$) and the thickness at the centre (d$_{\mathrm{c}}$).

The cylindrical edge of the disc is characterized by its temperature, T$_{d}$, and the conical surface of the disc by a radial temperature profile obtained by modifying the simple $\alpha$ temperature distribution. Assuming  that the disc is in physical and thermal contact with the gainer, so the inner radius and temperature of the disc are equal to the temperature and radius of the star (${\cal R}_{h}$, T$_{h}$). In this way, we  describe the temperature distribution by the following law (\cite{D2}):
\begin{equation}\label{model_disc}
T(r)=T_{d}+\big(T_{h}-T_{d}\big)\Big[1-\Big(\frac{r-R_{h}}{R_{d}-R_{h}}\Big)^{a_{T}}\Big]
\end{equation}
 The temperature of the disc at the edge (T$_{d}$) and the temperature exponent (a$_{\mathrm{T}}$), as well as the radii of the star (${\cal R}_{h}$) and of the disc (${\cal R}_{d}$) are free parameters, determined by solving the inverse problem.
 
 According to the temperatures of the system components, we set the
gravity-darkening exponents and component's albedos at their
theoretical values (${\rm\beta_{1,2}=0.25}$, ${\rm A_{1,2}=1.0}$),
corresponding to von Zeipel's law for fully radiating shells and
complete re-radiation.

Limb-darkening was applied by using the nonlinear approximation by \cite{cla00} for the photometric bands. The limb-darkening coefficients are interpolated in each iteration for the current values of the effective temperature, ${\rm T_{eff}}$, and surface gravity, $\log_{10} g$, as is described in detail by \citet{djur04}. The limb-darkening was applied to the disk in the same way, with
$\log_{10} g$ that corresponds to the middle of the disk radius.

 The models of the system can be refined by introducing active regions on the edge of the disc. The active regions have a higher local temperatures so their inclusion results in a non-uniform distribution of radiation. The existence of such regions (hot and bright spots) can be explained by the gas dynamic of interacting binaries, see e.g.  \cite{H2} and  Bisikalo et al. (1998, 2000, 2005).
 Based on these investigations of systems with accretion discs, we recently updated our model to allow inclusion of the hotspot (hs) and bright spot (bs) active regions.

These regions are characterized  by the following parameters: A$_{\mathrm{hs, bs}}=$T$_{\mathrm{hs,bs}}\,/\,$T$_{d}$ - the hot and bright spot to disc temperature ratios,  $\theta_{\mathrm{hs, bs}}$ and $\lambda_{\mathrm{hs, bs}}$ - the spot's angular dimensions and longitudes (in arc degrees) and $\theta_{\mathrm{rad}}$ - the angle between the line perpendicular to the local disc edge surface and the direction of the hot-spot maximum  radiation. The longitude $\lambda$  is measured clockwise (as viewed from the direction of the +Z-axis, which is orthogonal to the orbital plane) with respect to the line connecting the star centres (+X-axis), in the range $0-360$ degrees  (for more details, see \cite{D3}). These parameters are also determined by solving the inverse problem.
Including these active regions in the model, we found significant improvements in the quality of the fit to observations. For further information about  these active regions in the model,  see  \cite{D5}.

One possible limitation of the code at its present implementation
is the lack of a detailed treatment for the donor irradiation by the disk part
facing the donor, including hot spot. This effect could be potentially
important in close binaries with a large difference in stellar and
disk temperatures. However, it should be a second-order effect compared
with the stellar and disc flux contributions (including donor irradiation by the gainer) already implemented in our code.
This is demonstrated in the very good fit  obtained with the orbital multi-wavelength light curves.

\subsection{Fitting Procedure  for photometric data}\label{fitting_procedure}

The fitting was done for a limited set of  important parameters. The mass ratio $q={\cal M}_{c}/{\cal M}_{h}=0.21$ (where the subscripts $h$ and $c$ stand for the more massive hotter gainer and less massive cooler donor) was set to the value determined  by the radial velocities of our new FLAMES observations, discussed in Section \ref{spectroscopy}.  The
temperature for the primary was fixed to the value given by M08, viz.\, 25\,000 K. Rotation for the donor was assumed synchronous  ($f_{c}=1.0$)  since it is assumed that it has filled its Roche lobe (i.e. the filling factor of the donor was set to $F_{c}=1.0$). In the case of the gainer, however, the accreted material from the disc is expected to transfer enough angular moment to increase the spin rate of the gainer up to the critical rotation velocity as soon as even a small fraction of the mass has been transferred (\cite{Mink1}). This means that the gainer fills its corresponding non-synchronous Roche lobe for the star rotating in the critical regime.  This assumption is justified by the rapid spin-up of the gainer at the system evolutionary stage (see Section 4). So, the dimensions and the amount of rotational distortion are uniquely determined by the factor of non-synchronous rotation, which is the ratio between the actual and the Keplerian angular velocity. In this model the factor of non-synchronous rotation for the gainer was a free parameter also determined by solving the inverse problem. However, we also
calculated the best fit model for a synchronous gainer, and the results  practically do not differ from the critical case.
The results of our analysis are presented in Table 4.

The stellar parameters improve significantly the solution given by M08.
We find primary and donor  stars with 9.1 $\pm$ 0.5 $M_{\odot}$ and 1.9 $\pm$ 0.2 $M_{\odot}$, respectively. They have $\log\,g$ of 3.9 $\pm$ 0.1 and 2.8 $\pm$ 0.1 and radii of 5.6 $\pm$ 0.2 $R_{\odot}$ and 8.9 $\pm$ 0.3 $R_{\odot}$, respectively. The bolometric magnitudes are $M^{h}_{\mathrm{bol}} = -5.3$ and  $M^{c}_{\mathrm{bol}} = -3.4$. The best fit requires an optically thick disc around the gainer. This disc has a moderately convex shape, with central thickness  $\mathrm{d_{c}} \approx 2.6\, \mathrm{R}_{\odot}$  and thickness at the edge  $\mathrm{d_{e}} \approx 0.8\, \mathrm{R}_{\odot}$.  The radius of the disc is  ${\cal R}_{d}\approx 14\, R_{\odot}$,  more than twice larger than the gainer radius  and about $\sim77$\% of the Roche lobe radius. The temperature of the disc increases from T$_{d}=12\,600$ K at its edge to  T$_{h}=25\,000$ K in the inner radius, where it is in thermal and physical contact with the gainer.

 The two spots on the edge of the disc mentioned in Section 3.2  were found at longitudes 324 and 130 degrees with temperatures 12 and 18 \% higher than the disc edge temperature, respectively.
Whereas the hot spot represents the region of impact of the gas stream with the disc, the bright spot at $\lambda \approx$  130 degrees can be interpreted as a region where the disc significantly deviates from circular shape, or as a region where the stream material falls back to the opposite side of the disc, after being deflected by the impact. 
Due to the relatively small size and temperature difference compared
to the lateral side of the disk, the hot spot is not sufficient to
irradiate the face of the donor significantly. 
The bright
spot is located on the disk lateral side which is not able to
contribute to the irradiation of the donor.

In Fig. \ref{fOGLE-n-synch}   we show the comparison between the synthetic and the OGLE-I, MACHO-B and MACHO-R light curves.
We do not include errors bars since the scatter in the light curves probably gives a better representation of the true errors. 
In general the fit is pretty good,  but there is a slightly larger scatter in $O-C$ around secondary eclipse (I-band) and main eclipse (b and r bands).
This scatter is very minor and could represent a third light in the system that is not considered  in the
model and that is best visible during eclipses. A probable explanation for these larger residuals is the 3D structure of the hot spot,
including  material located at high latitudes, whose small  contribution to the total flux is detected only during eclipses. We show also the geometrical model of the system at quadratures.
The residuals do not depend on the long-cycle phase. This important result indicates that the orbital solution including the disc remains constant during the whole long cycle. Therefore, the disc represented in the model is not the cause for the long cycle.  The figure also illustrates the geometrically thin nature of the disc, its relative size and the positions of the hot and bright spots in the system. At quadratures, the disc contributes about 15\% to the total light of the system at the $I$-band, whereas at main eclipse the secondary contributes about 84\%.


\begin{figure*}
\centering
 \includegraphics[width=0.7\textwidth, height=0.5\textwidth]{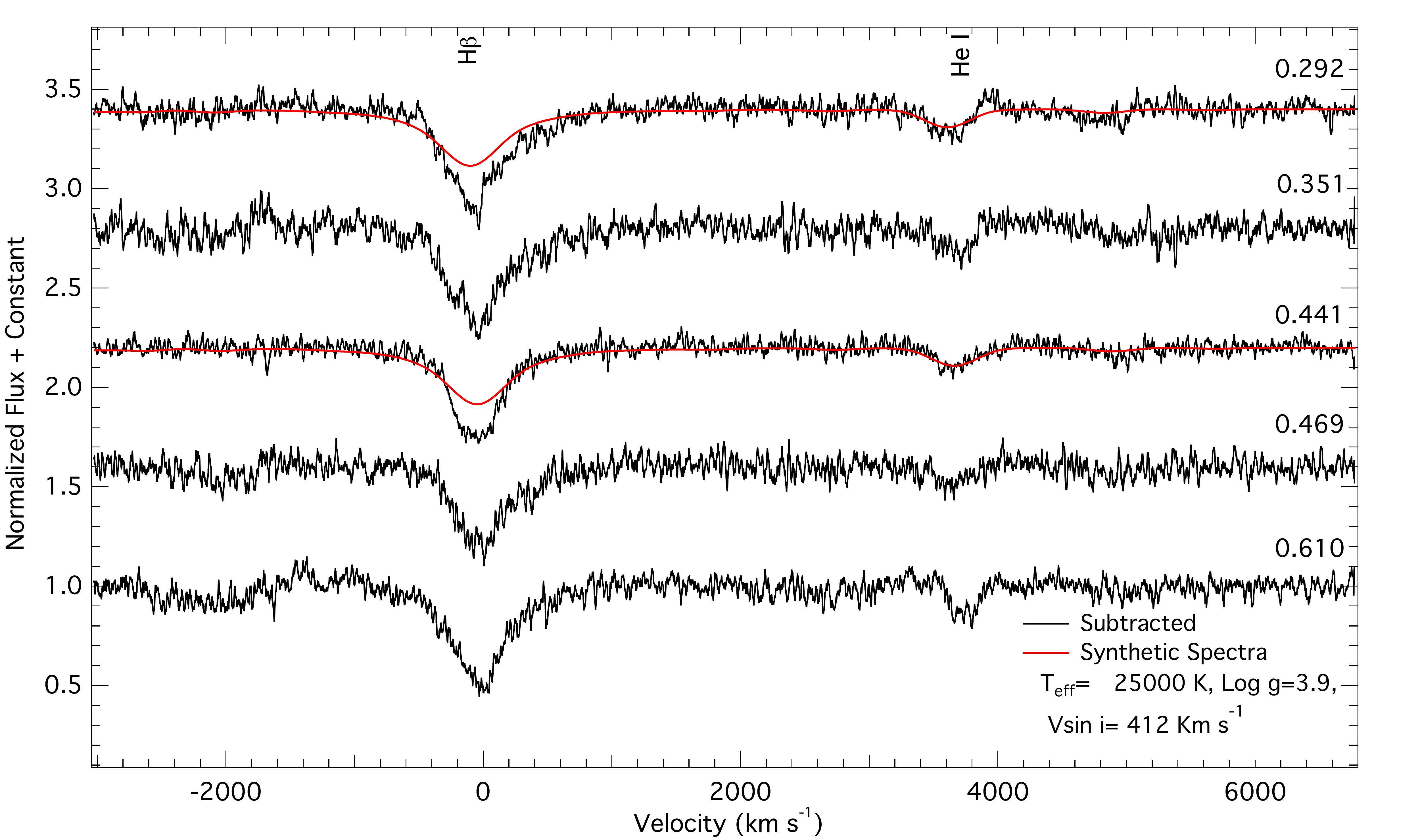}
\caption{Donor-subtracted spectra showing  H$_{\beta}$ variable extended wings up to $\pm 1000$ km s$^{-1}$. He I $4\,921\ \AA$  absorption line is also visible.  The red line overplotted on the donor-subtracted spectra is the SYNTHE model using Kurucz atmospheres  for a more-massive (gainer) component in critical rotation regime. The labels in the upper right of each spectrum indicate the orbital phase.}
 \label{donorsutracted}
\end{figure*}

\begin{figure*}
\centering
 \includegraphics[width=0.7\textwidth, height=0.5\textwidth]{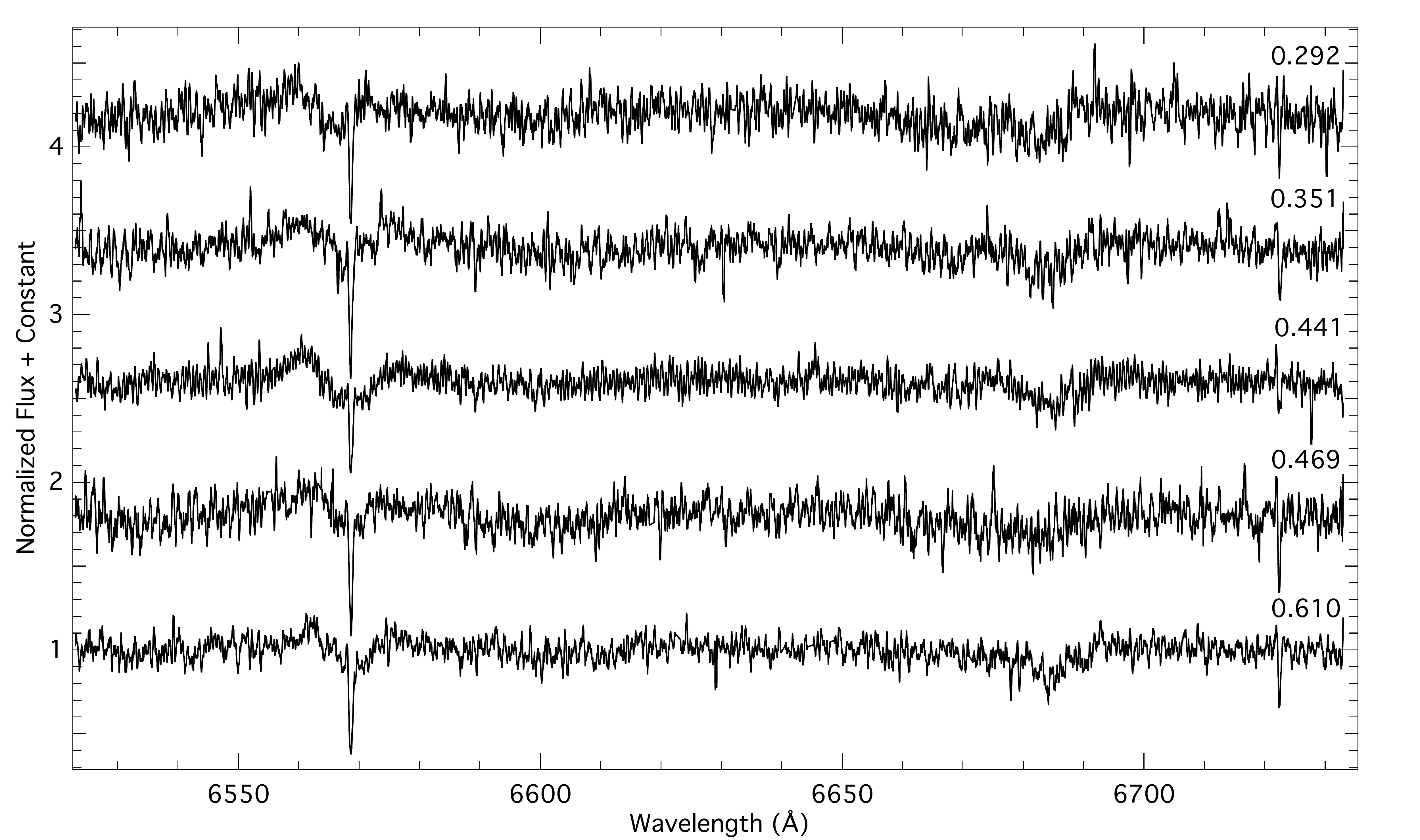}
\caption{Residual spectra showing the H$\alpha$ and He\,I 6\,678 line profiles  after removing the donor and gainer synthetic spectra. The narrow absorption in H$\alpha$ is an artifact of the reduction process (Section 2). The labels in the upper right of each spectrum indicate the orbital phase.}
 \label{Residual}
\end{figure*}

\subsection{Donor-subtracted and residual spectra}

One of the goals of this paper is to find information about the environment  in the vicinity of the hot primary star. However, from  the disentangling processes we know that our composite spectrum is contaminated by the light contribution of the B-type donor, in particular  in the Balmer and He lines. Thus, we need to remove the secondary star's spectral contribution from the observed spectra to study the light contribution of the primary  and its circumstellar gas.

In order to remove the donor light from the composite spectra we assume that its contribution to the total light is additive to the other light sources  and represented by the light-curve model proposed in the section \ref{light_curves_analysis}. Furthermore, we compute a synthetic spectrum using the data in Table 4 for the donor star.

The model for the donor stellar atmosphere was constructed for an effective temperature of $13\,000$ K and surface gravity of $\log g=3.0$ with solar abundance, using the LTE ATLAS9 code (\cite{K1}), which handles the line opacity with the opacity  distribution functions method (ODFs). The Kurucz models are constructed with the assumptions of plane-parallel geometry and hydrostatic and radiative equilibrium of the gas. The synthetic spectrum was computed with the SYNTHE code (\cite{K1})  and broadening at a rotational velocity of $V\mathrm{_{rot}\,sin}\,i=60$ km s$^{-1}$ , assuming that the donor has synchronous rotation, as was mentioned in Section 3.3. Both codes, ATLAS9 and SYNTHE were ported under  GNU Linux by \cite{Sb1} and are available online\footnote{wwwuser.oat.ts.astro.it/atmos/}. The atomic data were taken from \cite{Ca1}\footnote{wwwuser.oat.ts.astro.it/castelli/grids.html}.

\begin{figure}
    \centering
    \includegraphics[width=0.5\textwidth, height=0.35\textwidth]{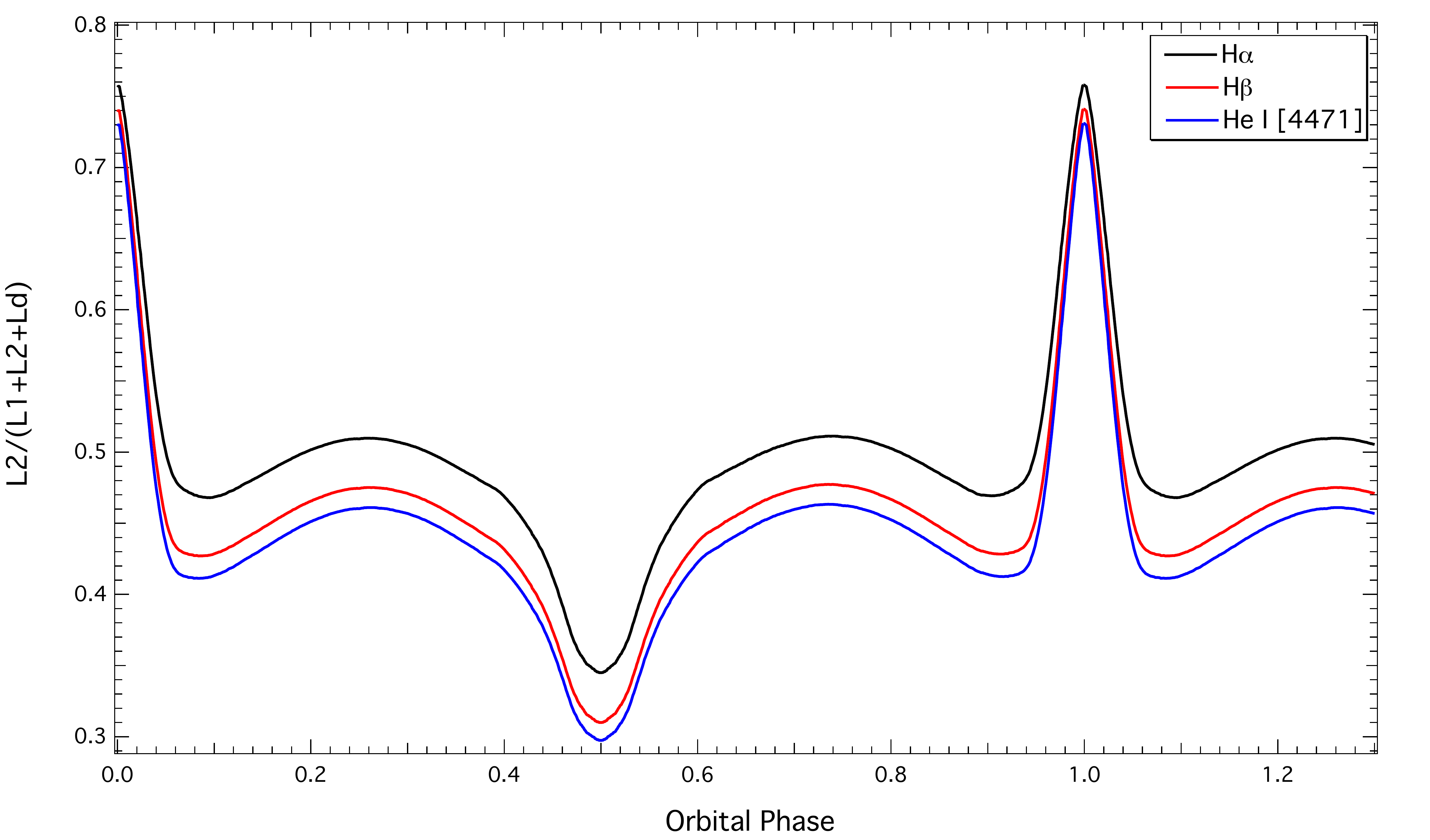}
    \caption{Light contribution factor for the donor at different spectral lines and orbital phases according to the light-curve model given in Section \ref{light_curves_analysis}. $L_{1}$, $L_{2}$ and $L_{d}$ are the gainer, donor and disc fluxes, respectively.}
    \label{light_contribution}
\end{figure}

After that, the synthetic spectrum was Doppler corrected and scaled according to the contribution of the donor star at a given orbital phase and at the given wavelength range. This method was used by \cite{M12b} to remove the donor spectrum from the observed composite spectrum of the interacting binary V393 Sco. Then the isolated gainer spectrum plus disc in units of its normalized flux is given by
\begin{equation}
f_{\mathrm{G}}\,(\lambda,\Phi_{0}, \Phi_{l})=f(\lambda,\Phi_{0}, \Phi_{l})-P\,(\Phi_{0},\lambda_{c})\times f_{\mathrm{D}}\,(\lambda,\Phi_{0})
\end{equation}
 where $f_{\mathrm{G}}$ is the donor-subtracted flux, $f$ the observed flux, $ f_{\mathrm{D}}$ the synthetic donor spectrum, $P$ the fractional contribution of the donor derived from our light-curve  model  and $\lambda_{c}$ the representative wavelength where $P$ is calculated. The theoretical light contribution factors for the donor at different spectral lines according to the light-curve  model are shown in Fig \ref{light_contribution}. The result  of this process was a set of  ``donor-subtracted" spectra that were normalized to the new continuum.

Our spectra cover  1.5 times the long cycle of 172 days, but only the Medusa spectra are well distributed to adequately map the long variability.
We inspected these profiles, having
subtracted the donor, but without observing  significant changes in the small wavelength range provided by the Medusa spectra. In the following, we analyze the UVES spectra,
that show the prominent  H$\beta$ and He\,I 4\,921\AA\  lines,  looking  for orbital
variability.

 The donor-subtracted spectra for the H$\beta$ and He I  4\,921 $\AA$ absorption lines in orbital phases 0.29 to 0.61 are shown in Fig.  \ref{donorsutracted}.
The H$\beta$ profiles are broad and show variable absorption wings extending up to $\pm 1000$ km s$^{-1}$. These lines are more symmetric near the secondary eclipse.
 The Helium lines are broad and asymmetric and sometimes show flanking emission, especially at the red side of the He\,I 4\,921 line in Fig.\,3.
Measures of Full Width at Half Maximum ($FWHM$) and equivalent width  ($EW$) for these lines are shown in Table 5 for each orbital phase.
Fig.\,3 also shows a synthetic spectrum computed with SYNTHE for the gainer  using parameters T$_{\mathrm{eff}}=25\,000$  K and $\log g=3.9$ and solar abundance.  Our synthesized spectrum was computed for a gainer in critical rotation using  a rotational parameter $V\mathrm{_{rot}\,sin}\,i=412$ km s$^{-1}$, as estimated in Section \ref{fitting_procedure}.

It is clear that the profiles are highly variable in shape. The fact that the H$\beta$ line is deeper than the gainer synthetic profile suggests that it is produced in an extended absorption media beyond the stellar photosphere, a kind of pseudo-photosphere. This media could be the disc revealed in the light curve analysis.  
The lack of obvious variability in the 13.7 year folded orbital light curve indicates a stable disc,
so the large profile variability suggests the presence of
additional mass streams in the system.
The general features are compatible with absorption/emission  in an extended photosphere around the gainer plus additional mass flows.

We found a good match using  the critical rotational velocity for the He I profiles and H$\beta$ absorption wings of our subtracted spectra, around secondary eclipse, when the gainer is in front to the donor, as shown in Fig.  \ref{donorsutracted}. However, we observe a small deviation at phase 0.29, when the wings of the line profiles are less symmetrical, showing extended absorption in the  H$\beta$ red wing and emission in the He\,4\,921\AA\  red wing.

We obtained  residual spectra around the H$\alpha$ region. These spectra were obtained from the donor-subtracted spectra after  removing the gainer synthetic spectrum. The process reveals weak double emission at H$\alpha$ with a peak separation of
about  620 km s$^{-1}$ (Fig.\,4). The same exercise at the He I 6\,678  $\AA$ line shows absorption and traces of emission.
We are cautious about interpreting the H$\alpha$ double emission as produced in the circumprimary disc, since in the case of V\,393 Scorpii the evidence indicates that this kind of profile is produced in a bipolar wind (Mennickent et al. 2012b).

We conclude that line profile fitting cannot be used as proof of critical rotation in this system due to the high variability of the profiles. On the other hand, the detection of Balmer and He\,I emission is consistent with the presence of circumstellar matter in the system.

 \begin{table}\label{FWHM_EW}
\begin{tabular}{cccccccc}
\hline
&\multicolumn{2}{c}{FWHM (km s$^{-1}$)} & &\multicolumn{2}{c}{EW (10$^{-2}$\AA)} \\
Phase & H$_{\beta}$ & He I [4921]& & H$\beta$ & He I [4921] \\
\hline
0.292 &$579.30\pm 3.60$  &$335.80\pm1.08 $  &  & $444\pm 1$   & $73\pm 1$    \\
0.351& $574.00\pm 1.19$ & $338.10\pm0.24 $  &  & $432\pm 1 $ & $63\pm 3$     \\
0.441&$462.53\pm1.05 $   &$301.73\pm4.15 $   &  &$364\pm 1 $ & $ 54\pm 2 $   \\
0.469& $ 523.13\pm 1.90$ &$ 247.00\pm 3.53$   &  &$333 \pm 1 $  &$ 40 \pm 1$   \\
0.610& $562.40 \pm 0.65 $ & $233.30 \pm 4.41$ &  &$442\pm 1$  &$ 58  \pm 1$  \\\hline
\end{tabular}
\caption{ Equivalent width and FWHM of the H$\beta$ and He  4\,921 $\AA$ absorption lines from the donor-subtracted UVES spectra.}
 \end{table}

 \begin{figure}
    \centering
    \includegraphics[width=0.5\textwidth, height=0.35\textwidth]{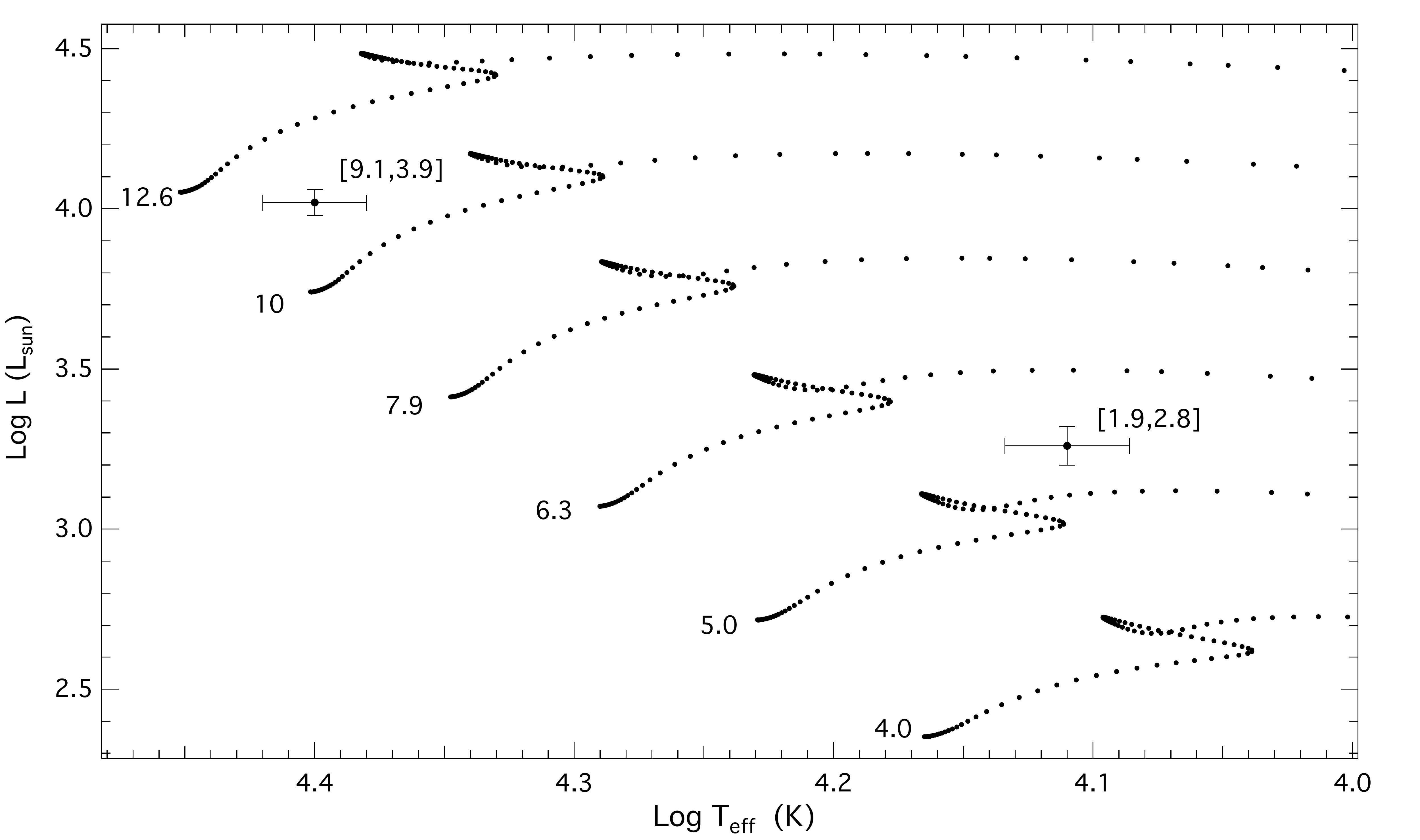}
    \caption{Comparison between evolutionary tracks for single stars with solar metallicity (Claret 2004) and the physical parameters of \va\  in the $\log \mathrm{T_{eff}}-$ $\log \mathrm{L}$ diagram. Derived mass and $\log g$ values are given between parentheses. The evolutionary tracks are labelled with the initial masses. }
    \label{single_star}
\end{figure}

 \begin{figure}
    \centering
    \includegraphics[width=0.5\textwidth, height=0.4\textwidth]{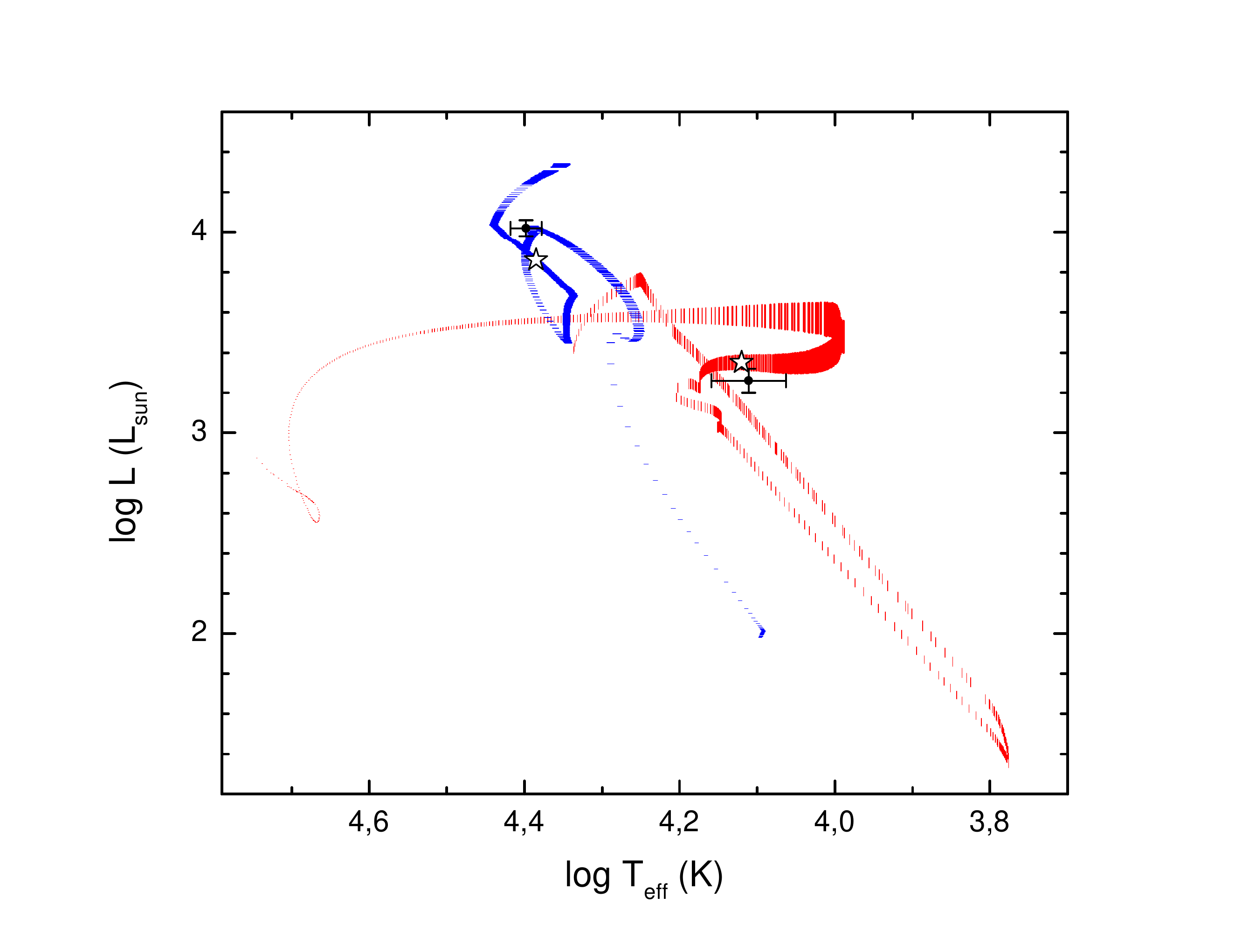}
    \caption{ Evolutionary tracks for the binary star model from Van Rensbergen et al. (2008) that best fit the data. Donor (red, vertical dashes track) and gainer (blue, horizontal dashes track) evolutionary paths are shown, along with the observations for \va (filled circles with errors bars). The best fit is reached at the time corresponding to the model indicated by stars, that is characterized in Table 6. Stellar sizes are proportional to the dash sizes.}
    \label{tracks}
\end{figure}

\section{Discussion}

\subsection{On the evolutionary stage of OGLE 05155332-6925581 }

The comparison  of the stellar  parameters of OGLE 05155332-6925581 with predictions of evolutionary models for single stars of solar metallicity
indicates that the gainer is located near the main-sequence band, a little displaced to higher luminosities for
a star of comparable mass, while the donor is  evolved from the main sequence, much more luminous for its mass, a fact explained by its expansion and subsequent filling of its Roche lobe (see Fig.\, \ref{single_star}). The small luminosity excess of the gainer can be explained in terms of their slightly advanced evolutionary stage; it has consumed about half of the hydrogen in its core, as will be shown later in this section.

In order to understand  the evolutionary stage of this binary  we compared our system parameters with those of the 561 conservative and non-conservative  evolutionary tracks given by \cite{V1}, available at the Center de Donn\'ees Stellaires (CDS). Similarly to the case of V393 Sco studied by  Mennickent et al. (2012a, hereafter M12a), a multi-parametric fit was carried out
for each synthetic model (labeled with $i$) through the evaluation of the quantity $\chi_{i,j}^{2}$ defined by:
\begin{equation}\label{chi_square}
\chi_{i,j}^{2}\equiv\frac{1}{N}\sum_{k}\omega_{k}\Biggl[\frac{S_{i,j,k}-O_{k}}{O_{k}}\Biggr]^{2}
\end{equation}
where $N$ is a normalization factor and $\omega_{k}$ the statistical weight of the parameter  $O_{k}$, calculated as:
\begin{equation}\label{weight}
\omega_{k}=\sqrt{O_{k}/\epsilon\,(O_{k})}
\end{equation}
where $\epsilon\,(O_{k})$ is the error associated with the observable $O_{k}$.  $S_{i,j,k}$ is the theoretical parameter for the observable $O_{k}$, in the time $t_{j}$ for the $i$-labeled model. For further information about the fitting procedure see M12a.

We find the absolute $\chi^{2}$  minimum in the tidal-strong interacting model with initial masses of 8  and 3.2 $M_{\odot}$ and initial orbital period $P_{\mathrm{orb}, i}= 2.5$ days. The absolute minimum $\chi_{\mathrm{min}}^{2}$ identifies the current age of the system along with the theoretical stellar and orbital parameters, which are presented in Table  6.

The best fit indicates that \va\  has an age of $4.76\times 10^{7}$ years. The corresponding evolutionary tracks for the primary and secondary stars are shown in Fig.  \ref{tracks}, along with the position for the best model for our system.

We notice a good agreement in the evolutionary tracks given by the best model. The secondary  star is inflated ($R_{c}= 9.1\, R_{\odot}$) and exchanges mass  through the L1 Lagrangian point at a rate of  $\dot{M}=3.105\times 10^{-6}\,M_{\odot}\, yr^{-1}$. This rather large mass transfer rate compares relatively well with the low limit found by  M08, viz.\, 8.6 $\times$ 10$^{-5}$ M$_{\odot}$/yr, derived assuming accretion powered luminosity.

\subsection{On mass flows and angular momentum loss }

The orbital light curve was re-examined with the program Period04\footnote{http://www.univie.ac.at/tops/period04/}. We calculated the error in the main frequency of the fit to the orbital light curve. The error consistently given by Monte Carlo simulations and the method of least squares is 2 $\times$ 10$^{-7}$ Hz. This means that the period could drift at most by 2 $\times$ 10$^{-5}$ d in 3365 d (the OGLE dataset time baseline). This implies that a period change at constant rate, if present, should be less than 0.2 s/yr to be consistent with the photometric time series. However, according to the best theoretical model, we find the system in a stage characterized by a rapid change of the orbital period (Fig.\,8). The period should be changing by 4.7 s/yr. This apparent inconsistency was already noted by M08, who argued that the effects of mass exchange and mass loss could be balanced in the system producing a net effect of non-variability for the orbital period.

The models by Van Rensbergen et al. (2008, 2011) parametrize mass and angular momentum loss from the system through the parameters $\beta$ and $\eta$, respectively.  The mass loss is driven by radiation pressure from a hot spot located in the stream impact region on the stellar surface or accretion disc edge. The mass loss extracts angular momentum from the system and in this particular model only from the gainer. According to the models of Van Rensbergen et al. (2011), particular pairs of  the parameters $\beta$ and $\eta$ should result in a constant orbital period.\\
The mass and angular momentum loss from a system during mass transfer can be described with the $\left(\beta,\eta\right)$-mechanism (see Rappaport et al. 1983). Here,
\begin{equation}
\beta = \left|\frac{\dot{M_h}}{\dot{M_c}}\right|
\end{equation}
is the mass transfer efficiency, i.e. the fraction of mass lost by the donor 
accreted by the gainer. 
The angular momentum loss $\dot{J}$ from the system resulting from a given amount of mass loss $\dot{M}$ is determined by $\eta$, according to

\begin{equation}
\dot{J}=\sqrt{\eta}~\dot{M}_{c}~(1-\beta)~\frac{2~\pi~a^{2}}{P},
\end{equation}
with $a$ the orbital seperation and $P$ the orbital period. It can be shown (see e.g. Podsiadlowski et al. 1992) that whenever $\beta$ and $\eta$ remain constant throughout the considered timeframe (with indices i for initial and f for final), the resulting period variation is given by \\
\\
for $0 < \beta < 1$:

\begin{equation}
\frac{P_\mathrm{f}}{P_\mathrm{i}} = \left(\frac{M_{c\mathrm{f}}+M_{h\mathrm{f}}}{M_{c\mathrm{i}}+M_{h\mathrm{i}}}\right)\left(\frac{M_{c\mathrm{f}}}{M_{c\mathrm{i}}}\right)^{3\left[\sqrt{\eta}\,\left(1-\beta\right)-1\right]}\left(\frac{M_{h\mathrm{f}}}{M_{h\mathrm{i}}}\right)^{-3\left[\sqrt{\eta}\frac{1-\beta}{\beta}+1\right]};
\end{equation}
\vspace{0.3cm}
for $\beta = 0$:

\begin{equation}
\frac{P_\mathrm{f}}{P_\mathrm{i}} = \left(\frac{M_{c\mathrm{f}}+M_{h\mathrm{f}}}{M_{c\mathrm{i}}+M_{h\mathrm{i}}}\right)\left(\frac{M_{c\mathrm{f}}}{M_{c\mathrm{i}}}\right)^{3\left(\sqrt{\eta}-1\right)}\mathrm{e}^{3\sqrt{\eta}\,\left(\frac{M_{c\mathrm{f}}-M_{c\mathrm{i}}}{M_{h\mathrm{i}}}\right)}.
\end{equation}
If it is assumed that mass is lost with the specific orbital angular momentum of the gainer, it can be shown that this yields

\begin{equation}
\eta = \left(\frac{M_{c}}{M_{c}+M_{h}}\right)^{4},
\end{equation}
resulting in $\eta\ll1$. On the other hand, if one assumes that matter is lost by the formation of a non-corotating circumbinary disc after passing through the second Lagrangian point, it was shown by Soberman et al. (1997) that a typical value is $\eta=2.3$.

 \begin{figure*}
    \centering
    \includegraphics[width=0.52\textwidth, height=0.45\textwidth]{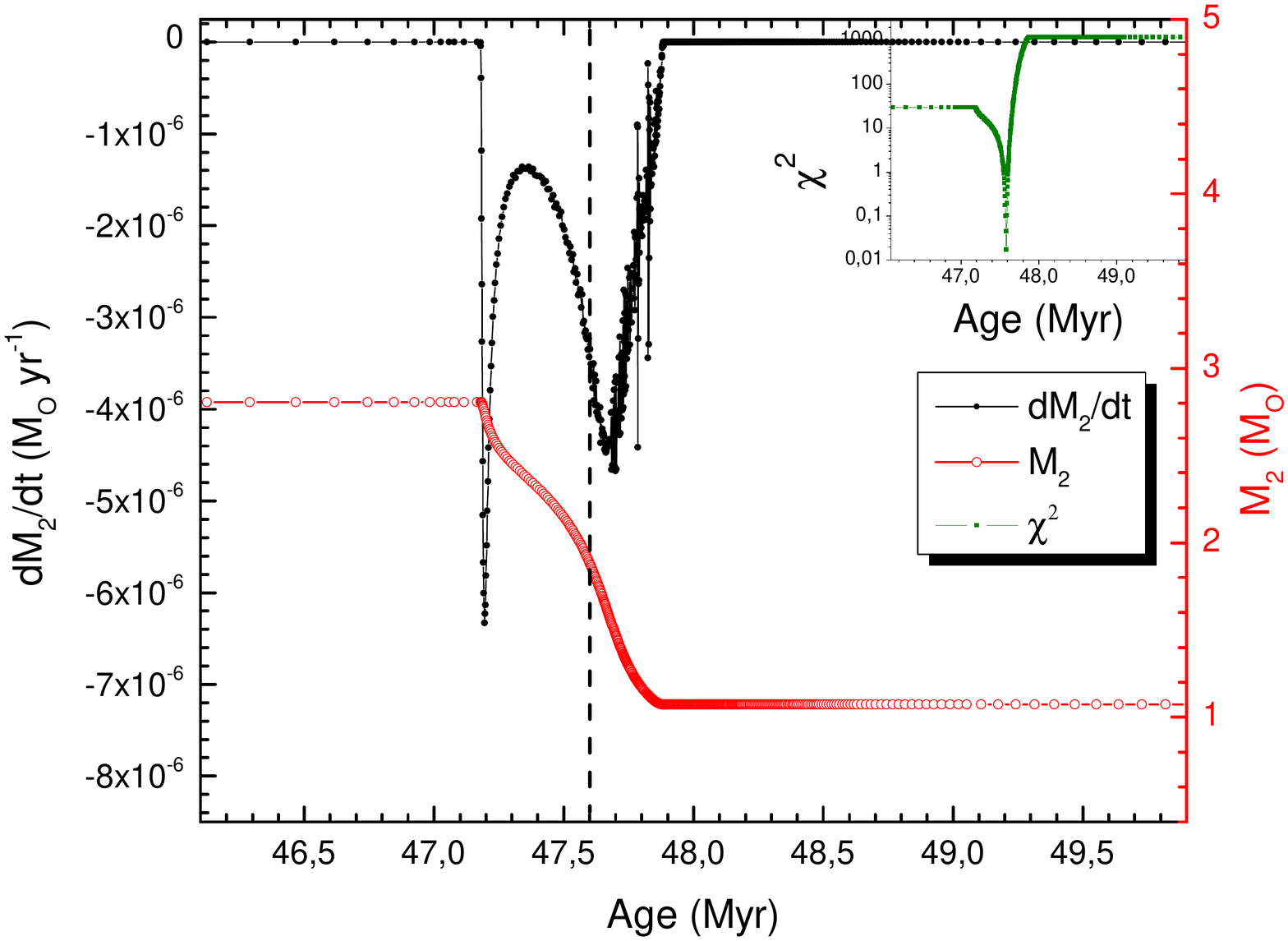}
     \includegraphics[width=0.47\textwidth, height=0.47\textwidth]{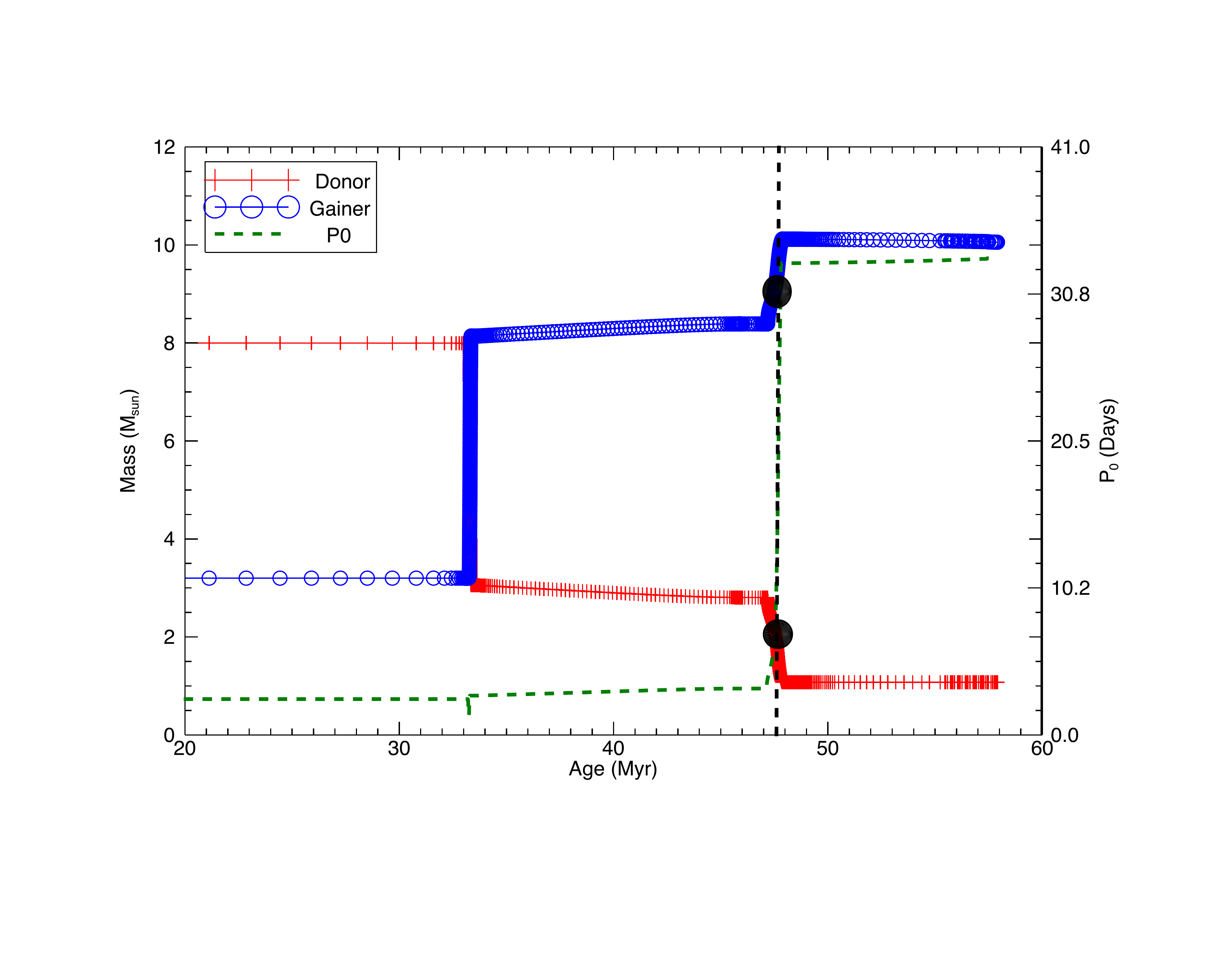}
    \caption{{\it Left}. $\dot{M}_{2}$ (upper curve) and $M_{2}$  for the best evolutionary model, $\chi^{2}$ is shown in the inset graph. The few spikes in the $\dot{M}_{2}$ curve reflect minor convergence artifacts produced during the numerical calculations. {\it Right}. Evolution of orbital period and mass of the components with the mass transfer time for \va\  with the initial orbital period P$_{\mathrm{orb},i}=2.5$ days.  The vertical dashed lines and filled circles indicates the position for the best model.}
    \label{evolutionary}
\end{figure*}

\begin{figure}
    \centering
    \includegraphics[width=0.5\textwidth, height=0.35\textwidth]{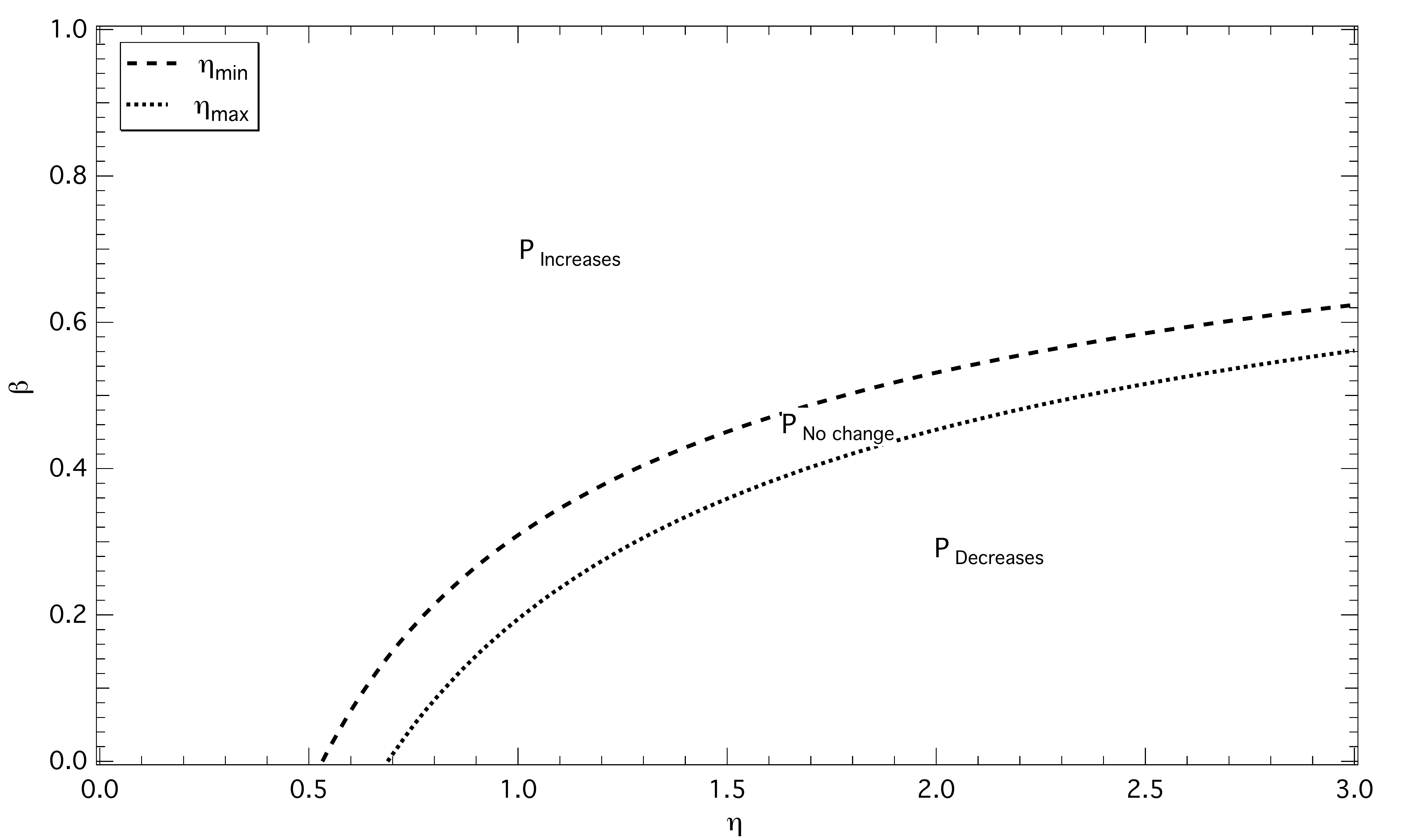}
    \caption{$\eta$ and $\beta$ parameters controlling the mass and angular momentum loss from the binary according to the model proposed by Rensbergen et al. (2011).
The space between the upper and lower curves is for systems with a constant orbital period, according to the observed boundaries for the orbital period variability. The region above the upper curve is for systems showing an
orbital period increase, while that below the lower curve is for systems with decreasing orbital period. The $\eta$ value for \va\  of 8.9 $\times$ 10$^{-4}$ is incompatible with a constant orbital period under the view of the aforementioned model.
}
    \label{OGLE_deltaP}
\end{figure}

Given the observational upper limit for the period variation in this paper, we have calculated the maximum amount by which the period may have changed during those 15 years. Then, with the equations for the period variation during (non-)conservative mass transfer and using the physical parameters of this particular system, we have calculated that 15 year period variation theoretically for all different combinations of $\beta$  and $\eta$. Results are shown in Fig \ref{OGLE_deltaP}.

The zone between the two curves is where the orbital period does not change more in 15 years than is allowed by the observations. The zone above the upper curve is where the period increases too much. This is the case for large $\beta$ (little mass loss) and/or small $\eta$ (little angular momentum loss). The zone below the lower curve is where the period decreases too much. This is the case for small $\beta$ (much mass loss) and large $\eta$ (much angular momentum loss).

The assumption of specific gainer orbital angular momentum loss (the one used in the calculation of the models) yields $\eta$ = 0.00089 for this system. It is obvious that this is way too low in order to fall into the constant period region, and that under this assumption, irrespective of $\beta$, the period will increase (much) more than allowed by the observations.
Based on this result we argue that another source of angular momentum loss is present in the system, possibly the outflows through $L_{2}$  and $L_{3}$  reported by M08. Notice that for $\eta \approx 2.3$,  representative for angular momentum loss from a circumbinary disc  (Soberman et al. 1997),
there is a mass loss ($\beta$ $\approx$  0.5) compatible with period constancy in this system.

 \begin{table}\label{Table_4}
\caption{The parameters of the Van Rensbergen et al. (2008) model that best fits the \va data. The hydrogen and helium core mass fractions ($X_{c}$ and $Y_{c}$) are given for the cool and hot star.}
\begin{tabular}{llll}
\hline
Quantity & Value & Quantity & Value \\\hline
Age & $4.76\times 10^{7}$\,yr & Period & 7.241\,d \\
$M_{c}$ & 1.945 $M_{\odot}$& $M_{h}$ & 9.249 $M_{\odot}$\\
$\dot{M}_{c}$ & $-3.17\times 10^{-6}\,M_{\odot}\, yr^{-1}$ & $\dot{M}_{h}$ & $3.105\times 10^{-6}\,M_{\odot}\, yr^{-1}$ \\
$\log T_{c}$ & 4.120 K &  $\log T_{h}$ & 4.385 K \\
$\log L_{c}$ & 3.351 $\,L_{\odot}$ &  $\log L_{h}$ & 3.862 $\,L_{\odot}$ \\
$R_{c}$ & 9.090$\,R_{\odot}$ &  $R_{h}$ & 4.821 $\,R_{\odot}$ \\
$X_{cc}$ & 0.00 & $X_{ch}$ & 0.496\\
$Y_{cc}$ & 0.98 & $Y_{ch}$ & 0.484\\\hline
\end{tabular}
\end{table}

\section{Conclusions}

We have presented a detailed spectroscopic and photometric study of \va\ including  the analysis of high resolution spectra and the application of a sophisticated light curve model. The multi-band light curves covering 13.7 years of photometric data from the MACHO and OGLE projects were analyzed along with the radial velocities of both stellar components of this eclipsing binary.

We find that the system is best modeled with a geometrically thin and optically thick disc around the gainer. The analysis of the photometric data has allowed  us to derive improved orbital parameters and physical stellar properties such as stellar masses, radii, luminosities and the effective temperatures. All these parameters are given in Table 4. Interestingly, the orbital solution remains constant during the long cycle. This means that the disc is not the origin of the long cycle, which is also consistent with previous claims that the source of this variability is not eclipsed (M08). On the other hand, the presence of variable Balmer emission suggests the existence of a fourth component in the system with physical conditions usually associated to optically thin circumstellar gas. Due to the presence of this disturbing component, our line profile investigation is not conclusive regarding gainer critical rotation.

Taking advantage of our new and improved system and stellar parameters, we explored the evolutionary stage of \va\, with the aid of published grids of evolutionary routes for
binary systems of similar masses, considering conservative and non-conservative evolution. In particular, the grid of models by Rensbergen et al. were considered in our work.
 We find the system in a semi-detached configuration and at a stage of rapid mass transfer. The donor has exhausted its hydrogen in its core and we observed the systems in a Case-B mass transfer, with an age of 4.76 $\times$  10$^{7}$ years. Contrary to the case of V\,393 Scorpii, the gainer parameters do not deviate too much from a main sequence star of similar mass.
This is consistent with a younger accreting object, in the sense that \va\ has not had time to accrete large amounts of matter since it is still inside the burst of mass transfer (Fig.\,8).
On the contrary,  V\,393 Scorpii already passed this event in the M12a model and had time to accumulate an important amount of mass in a massive disc.
Our studies indicate that the DPV phenomenon is observed at different evolutionary stages; Case-A for V\,393 Sco (M12a) and DQ\,Vel (Barria et al., in preparation) and Case-B for \va\ but always inside or after
a main mass transfer burst.

One notable observation in \va\ is the absence of orbital period change, overall considering the relatively large mass transfer rate. This cannot be explained in terms of the mechanism of mass and angular momentum loss proposed by Rensbergen et al. (2008, 2011). Actually, the best model indicates that the system should be found at a phase of  fast period increase. We argue that mass outflows through the Lagrange $L_{2}$ and $L_{3}$ points (as claimed by M08 to explain infrared spectroscopic observations) could extract angular momentum from the system in order to balance the mass exchange and produce a relatively constant orbital period. In addition, if the Balmer and He\,I emissions are due to a bipolar wind as suggested in the case of the DPV V\,393 Scorpii (M12b), this wind could be an additional channel for escape of angular momentum from the system.

It is worthy to mention that current theoretical evolutionary models for close binaries do not include neither predict the DPV phenomenon.
As the number of DPVs is relatively large, the phenomenon corresponds to a relatively long phase in the life of close binaries of intermediate mass. Our study of \va\ suggests that the DPV phenomenon could have an important effect in the balance of mass and angular momentum during the system evolution. This makes sense if the long cycle in \va\ turns out to be due to a cyclic bipolar wind as suggested for the DPV V\,393 Scorpii by M12b.


\section{Acknowledgments}

We thank an anonymous referee for the useful criticism of a first 
version of this paper. H.G. acknowledges his thesis co-advisers Christophe Martayan and Maja Vu{\v c}kovi\'c.
R.E.M. acknowledges support by Fondecyt grant 1110347 and  from the BASAL
Centro de Astrof\'isica y Tecnologias Afines (CATA) PFB--06/2007.
G.D. gratefully acknowledge the financial support of the Ministry of Education and Science of the Republic of Serbia through the project 176004, ÓStellar physicsÓ.


\begin{thebibliography}{}

\bibitem[\protect\citeauthoryear{Alcock et al., 1993}{}]{A1} Alcock C., et al., 1993, Nature, 365, 621.

\bibitem[\protect\citeauthoryear{Bisikalo et al.}{1998}]{B1} Bisikalo D. V., Boyarchuck A. A., Chechetkin V. M., Kuznetsov O. A., Molteni D., 1998, MNRAS, 300, 39

\bibitem[\protect\citeauthoryear{Bisikalo et al.}{2000}]{B2} Bisikalo D. V.,  Harmanec P., Boyarchuck A. A.,  Kuznetsov O. A., Hadrava P., 2000, A\&A, 353, 1009

\bibitem[\protect\citeauthoryear{Bisikalo et al.}{2005}]{B3} Bisikalo D. V.,  Kaigorodov P. V., Boyarchuck A. A.,  Kuznetsov O. A.,  2005, Astron. Rep., 49, 701

\bibitem[\protect\citeauthoryear{Castelli \& Hubrig}{2004}]{Ca1} Castelli, F. \& Hubrig, S., 2004, A\&A, 425, 263.


\bibitem[\protect\citeauthoryear{Claret}{2000}]{cla00} Claret A., 2000, A\&A 363, 1081


\bibitem[\protect\citeauthoryear{Claret 2004}{}]{C1} Claret A., 2004, A\&A, 424, 919.


\bibitem[\protect\citeauthoryear{de Mink , Pols \& Glebbeek 2007}{}]{Mink1} de Mink S. E., Pols O. R., Glebbeek E., 2007, in Stancliffe R. J., Houdek G., Martin R. G., Tout C. A., eds, AIP Conf. Ser. Vol. 948, Unsolved Problems in Stellar Physics. Am Inst. Phys., New York, p.321

\bibitem[\protect\citeauthoryear{Djura{\v s}evi\'c 1992}{}]{D1} Djura{\v s}evi\'c, 1992, Ap\&SS, 196, 267.

\bibitem[\protect \citeauthoryear{Djura\v sevi\'c et al. (2004)}{}]{djur04} Djura\v sevi\'c G., Albayrak B., Selam S. O., Erkapi\'c S., Senavci H.V.,  2004, NewA, 9, 425

\bibitem[\protect\citeauthoryear{Djura{\v s}evi\'c et al. 2008}{}]{D2} Djura{\v s}evi\'c G., Vince I., Atanackovi\'c O.,  2008, AJ, 136, 767.

\bibitem[\protect\citeauthoryear{Djura{\v s}evi\'c et al., 2010}{}]{D3}  Djura{\v s}evi\'c G.,  Latkovi\'c O.,  Vince I., Cs\'eki  A. , 2010, MNRAS, 409 ,329.

\bibitem[\protect\citeauthoryear{Djura{\v s}evi{\'c} et
al.}{2011}]{2011A&A...525A..66D} Djura{\v s}evi{\'c} G., Y{\i}lmaz M., Ba{\c s}t{\"u}rk {\"O}., K{\i}l{\i}{\c c}o{\u g}lu T., Latkovi{\'c} O., {\c C}al{\i}{\c s}kan {\c S}., 2011, A\&A, 525, A66

\bibitem[\protect\citeauthoryear{Djura{\v s}evi{\'c} et al. }{2012}]{D5} Djura{\v s}evi{\'c} G., Vince I., Antokhin I. I., Shatsky N. I., Cs{\'e}ki A., Zakirov M., Eshankulova M., 2012,
MNRAS, 420, 3081



\bibitem[\protect\citeauthoryear{Evans et al. (2011)}{}]{E1} Evans C. J.,  Taylor W. D., H\'enault-Brunet V., et al., 2011, A\&A, 530, A108



\bibitem[\protect\citeauthoryear{Heemskerk}{1994}]{H2} Heemskerk M. H. M., 1994, A\&A, 288, 807

\bibitem[\protect\citeauthoryear{Hilditch}{2001}]{2001icbs.book.....H} Hilditch R.~W., 2001, An Introduction to Close Binary Stars, Cambridge University Press.

\bibitem[\protect\citeauthoryear{Kurucz, 1993}{}]{K1} Kurucz R., 1993, CD-ROM 18

\bibitem[\protect\citeauthoryear{Li and Zhang }{2006}]{Li} Li L. \& Zhang F., 2006, New Astron, 11, 588


\bibitem[\protect\citeauthoryear{Mennickent et al.}{2003}]{M03} Mennickent R. E., Pietrzy\'nski G., Diaz M.,  Gieren W., 2003, A\&A,  399, L47

\bibitem[\protect\citeauthoryear{Mennickent et al.}{2005}]{M05a} Mennickent R. E., Cidale L., Diaz M., Pietrzy\'nski G., Gieren W., Sabogal B., 2005a, MNRAS,  357, 1219

\bibitem[\protect\citeauthoryear{Mennickent et al.}{2005}]{M05b} Mennickent R. E., Assmann P., Pietrzy\'nski G., Gieren W., 2005b,in Sterken C., ed., ASP Conf. Ser. Vol., 335, The Light-Time Effect in AStrophysics. Astron. Soc. Pac., San Francisco, p. 129



\bibitem[\protect\citeauthoryear{Mennickent et al.}{2008}]{M1} Mennickent R.E.,  Kolaczkowaki Z., Michalska G., et al., 2008, MNRAS, 389, 1605.

\bibitem[\protect\citeauthoryear{Mennickent et al., 2012}{}]{M12a} Mennickent R.E.,  Djurasevic D., Kolaczkowski Z. and Michalska G., 2012, MNRAS, 421, 862. (M12a)

\bibitem[\protect\citeauthoryear{Mennickent et al.}{2012b}]{M12b} Mennickent R.E., Kolaczkowaki Z., Djurasevic G., Niemczura E., et al., MNRAS in press. (M12b)







\bibitem[\protect\citeauthoryear{Podsiadlowski et al. 1992}{}]{Po1} Podsiadlowski P., Joss P.,  \& Hsu J. 1992, ApJ, 391, 246


\bibitem[\protect\citeauthoryear{Poleski et al. 2010}{}]{P1} Poleski  R. ,  Soszy\'nski I., Udalski A. , Szyma\'nski M. K. ,  Kubiak M., Pietrzy\'nski  G.,  Wyrzykowski L.,  Ulaczyk K., 2010, Acta Astron., 60, 179.


\bibitem[\protect\citeauthoryear{Rappaport et al. 1983}{}]{R1} Rappaport S., Verbunt, F., \& Joss, P., 1983, ApJ, 275, 713

\bibitem[\protect\citeauthoryear{Soberman et al.  }{1997}]{So} Soberman G., Phinney E., \& Van den Heuvel E., 1997, A\&A, 327, 620

\bibitem[\protect\citeauthoryear{Sbordone et al.,}{2005}]{Sb1} Sbordone, L.,  2005, MSAIS, 8, 61


\bibitem[\protect\citeauthoryear{Szyma\'nski 2005}{}]{S1}  Szyma\'nski M., 2005, Acta Astron., 55, 43.


\bibitem[\protect\citeauthoryear{Udalski et al., 1997}{}]{U1} Udalski A., Kubiak M., Szyma\'nski M., 1997, Acta Astron., 47, 319

\bibitem[\protect\citeauthoryear{Van Rensbergen et al.}{2008}]{2008A&A...487.1129V} Van Rensbergen W., De Greve J. P., De Loore C., Mennekens N., 2008, A\&A, 487, 1129

\bibitem[\protect\citeauthoryear{Van Rensbergen et al.}{2008}]{V1} Van Rensbergen W., De Greve J. P., De Loore C., Mennekens N., 2008, VizieR On-line Data Catalog, 348, 71129.

\bibitem[\protect\citeauthoryear{Van Rensbergen et al.}{2011}]{2011A&A...528A..16V} Van Rensbergen W., de Greve J. P., Mennekens N., Jansen K., de Loore C., 2011, A\&A, 528, A16


\bibitem[\protect\citeauthoryear{Wade 
\& Horne}{1988}]{1988ApJ...324..411W} Wade R.~A., Horne K., 1988, ApJ, 324, 411 

\bibitem[\protect\citeauthoryear{Walborn et al. 2002}{}]{W1} Walborn N. R., Ma\'iz-Apell\'aniz J., Barb\'a R. H., 2002, AJ, 124, 1601.


\bibitem[\protect\citeauthoryear{Watson et al.}{2003}]{2003MNRAS.341..129W} 
Watson C.~A., Dhillon V.~S., Rutten R.~G.~M., Schwope A.~D., 2003, MNRAS, 
341, 129 


\bibitem[\protect\citeauthoryear{Er-Gang et al. }{2010}]{Ze1} Zhao Er-Gang, Qian, Sheng-Bang,  E. Fern\'andez Lajœs, C. von Essen, Li-Ying Zhu, 2010, RAA, 10 438.





\end{thebibliography}
\end{document}